\documentclass[12pt,fleqn,a4paper]{article} 
\usepackage{epsfig,graphics,graphicx}
\usepackage{clshan-math,clshan-dm-dd}
\def \lsim {\:\raisebox{-0.7ex}{$\stackrel{\textstyle<}{\sim}$}\:}
\def \gsim {\:\raisebox{-0.7ex}{$\stackrel{\textstyle>}{\sim}$}\:}
\begin{document}
\thispagestyle{empty}
\begin{flushright}
 March 2011
\end{flushright}
\begin{center}
{\Large\bf
 Estimating the Spin--Independent WIMP--Nucleon   \\ \vspace{0.2cm}
 Coupling from Direct Dark Matter Detection Data} \\
\vspace{0.7cm}
 {\sc Chung-Lin Shan} \\
\vspace{0.5cm}
 {\it Department of Physics, National Cheng Kung University \\
      No.~1, University Road,
      Tainan City 70101, Taiwan, R.O.C.}                    \\~\\
 {\it Physics Division,
      National Center for Theoretical Sciences              \\
      No.~101, Sec.~2, Kuang-Fu Road,
      Hsinchu City 30013, Taiwan, R.O.C.}                   \\~\\
 {\it E-mail:} {\tt clshan@mail.ncku.edu.tw}                \\
\end{center}
\vspace{1cm}
\begin{abstract}
 Weakly Interacting Massive Particles (WIMPs) are
 one of the leading candidates for Dark Matter.
 For understanding the nature of WIMPs and
 identifying them among new particles produced at colliders
 (hopefully in the near future),
 determinations of their mass and couplings on nucleons
 from direct Dark Matter detection experiments are essential.
 Based on our model--independent method
 for determining the WIMP mass
 from experimental data,
 I present a way to also estimate
 the spin--independent (SI) WIMP--nucleon coupling
 by using measured recoil energies directly.
 This method is independent of
 the velocity distribution of halo WIMPs
 as well as (practically) of the as yet unknown WIMP mass.
 In a background--free environment,
 for a WIMP mass of $\sim$ 100 GeV
 the SI WIMP--nucleon coupling
 could in principle be estimated
 with an uncertainty of $\sim$ 15\%
 by using \mbox{2 (or 3) $\times$ 50 events} from experiments.
\end{abstract}
\clearpage
\section{Introduction}
 Astronomical observations and measurements indicate that
 more than 80\% of all matter in the Universe is dark
 (i.e., interacts at most very weakly
  with electromagnetic radiation and ordinary matter).
 The dominant component of this cosmological Dark Matter
 must be due to some yet to be discovered, non--baryonic particles.
 Weakly Interacting Massive Particles (WIMPs) $\chi$
 arising in several extensions of
 the Standard Model of electroweak interactions
 are one of the leading candidates for Dark Matter.
 WIMPs are stable particles
 with masses roughly between 10 GeV and a few TeV
 and interact with ordinary matter only weakly
 (for reviews, see Refs.~\cite{SUSYDM96, Bertone05}).

 Currently,
 the most promising method to detect different WIMP candidates
 is the direct detection of the recoil energy
 deposited in a low--background underground detector
 by elastic scattering of ambient WIMPs off target nuclei
 \cite{Smith90, Lewin96}.
 The recoil energy spectrum can be calculated
 from an integral over the one--dimensional
 velocity distribution function of halo WIMPs, $f_1(v)$,
 where $v$ is the absolute value of the WIMP velocity
 in the laboratory frame.
 In our earlier work \cite{DMDDf1v},
 we presented a way to reconstruct
 this one--dimensional velocity distribution function
 and to estimate its moments
 from the recoil spectrum as well as
 from measured recoil energies {\em directly}
 in direct Dark Matter detection experiments.
 {\em Neither} the WIMP--nucleus scattering cross section
 {\em nor} the local WIMP density
 is required in this analysis.

 However,
 the mass of halo WIMPs is needed for
 the reconstruction of the (moments of the)
 WIMP velocity distribution.
 Therefore,
 as the next step
 we developed a model--independent method
 based on the reconstruction of the moments of $f_1(v)$
 for determining the WIMP mass $\mchi$
 by combining two sets of (future) experimental data
 with different target nuclei directly
 \cite{DMDDmchi-SUSY07, DMDDmchi}.
 To do so,
 one simply requires that
 the values of a given moment of $f_1(v)$
 estimated by both experiments agree.
 This leads to a simple expression for determining $\mchi$,
 which can be solved analytically
 and each moment can be used.
 Moreover,
 by assuming that
 the ratio of the spin--independent (SI) scattering
 cross sections on protons and on neutrons is known,
 an additional expression for determining $\mchi$ has be derived.
 By combining the estimators for different moments with each other
 and with the estimator derived
 by making the assumption about
 the ratio of the SI cross sections,
 one can yield the best estimate of the WIMP mass
 \cite{DMDDmchi}.
 Here we found again that
 neither a prior knowledge about the WIMP--nucleus cross section
 nor that about the local WIMP density
 is required.

 Meanwhile,
 in the second method for the determination of the WIMP mass,
 the product of the local WIMP density times
 the SI WIMP--proton cross section,
 $\rho_0 \sigmapSI$,
 appearing in the expression for the scattering spectrum
 cancels out when we use the identity of this product
 for two different targets.
 Hence,
 as will be shown in the paper,
 once the WIMP mass can be determined
 one could then use this information
 to estimate $\sigmapSI$ conversely.
 Remind that,
 in order to identify new particles produced at e.g.,
 the Large Hardon Collider (LHC)
 to be indeed WIMPs detected by direct detection \cite{Baer},
 estimates of or constraints on their mass and couplings on nucleons
 from direct detection experiments are essential.
 However,
 due to the degeneracy between $\rho_0$ and $\sigmapSI$,
 for estimating the SI WIMP cross section by this method
 one has to make an assumption for the local WIMP density,
 which can so far be estimated
 with an uncertainty of a factor of $\sim$ 2
 \cite{SUSYDM96, Bertone05}.
 Nevertheless,
 our simulations show that,
 in spite of the large statistical uncertainty
 due to very few events,
 for a WIMP mass of $\sim$ 100 GeV,
 $\sigmapSI$ could be estimated
 with an uncertainty of 30\%
 by using 2 (or 3) $\times$ 50 events from experiments.
 This result is (much) better than
 our estimate of the local Dark Matter density.

 The remainder of this article is organized as follows.
 In Sec.~2
 I discuss the possibility
 of constraining the WIMP mass and its coupling on nucleons
 from a single experiment.
 In Sec.~3
 I present the method for estimating
 the spin--independent WIMP--nucleon coupling
 by combining two (or more) experiments.
 Some numerical results
 based on Monte Carlo simulations of future experiments
 will also be presented.
 In Sec.~4 the analysis will be extended to the case
 of spin--dependent (SD) WIMP--nucleon couplings.
 I conclude in Sec.~5.
 Some technical details for our analysis
 will be given in an appendix.
\section{Constraining the SI WIMP--nucleon coupling}
 The basic expression for the differential event rate
 for elastic WIMP--nucleus scattering is given by \cite{SUSYDM96}:
\beq
   \dRdQ
 = \calA \FQ \int_{\vmin}^{\vmax} \bfrac{f_1(v)}{v} dv
\~.
\label{eqn:dRdQ}
\eeq
 Here $R$ is the direct detection event rate,
 i.e., the number of events
 per unit time and unit mass of detector material,
 $Q$ is the energy deposited in the detector,
 $F(Q)$ is the elastic nuclear form factor,
 $f_1(v)$ is the one--dimensional velocity distribution function
 of the WIMPs impinging on the detector,
 $v$ is the absolute value of the WIMP velocity
 in the laboratory frame.
 The constant coefficient $\calA$ is defined as
\beq
        \calA
 \equiv \frac{\rho_0 \sigma_0}{2 \mchi \mrN^2}
\~,
\label{eqn:calA}
\eeq
 where $\rho_0$ is the WIMP density near the Earth
 and $\sigma_0$ is the total cross section
 ignoring the form factor suppression.
 The reduced mass $\mrN$ is defined by
\beq
        \mrN
 \equiv \frac{\mchi \mN}{\mchi + \mN}
\~,
\label{eqn:mrN}
\eeq
 where $\mchi$ is the WIMP mass and
 $\mN$ that of the target nucleus.
 Finally,
 $\vmin$ is the minimal incoming velocity of incident WIMPs
 that can deposit the energy $Q$ in the detector:
\beq
   \vmin
 = \alpha \sqrt{Q}
\~,
\label{eqn:vmin}
\eeq
 with the transformation constant
\beq
        \alpha
 \equiv \sfrac{\mN}{2 \mrN^2}
\~,
\label{eqn:alpha}
\eeq
 and $\vmax$ is the maximal WIMP velocity
 in the Earth's reference frame,
 which is related to
 the escape velocity from our Galaxy
 at the position of the Solar system,
 $\vesc~\gsim~600$ km/s.

 The local WIMP density at the position of the Solar system,
 $\rho_0$,
 appearing in the expression (\ref{eqn:dRdQ})
 for the scattering event rate
 has conventionally been determined
 by means of the measurement of
 the rotation curve of our Galaxy.
 The currently most commonly used value for $\rho_0$ is
 \cite{SUSYDM96, Bertone05}
\beq
         \rho_0
 \approx 0.3~{\rm GeV / cm^3}
\~.
\label{eqn:rho0}
\eeq
 However,
 as mentioned in the introduction,
 due to our location inside the Milky Way,
 it is more difficult to measure
 the accurate rotation curve of our own Galaxy
 than those of other galaxies;
 an uncertainty of a factor of $\sim$ 2
 has thus usually been adopted \cite{SUSYDM96, Bertone05}%
\footnote{
 Recently,
 some new techniques have been developed
 for determining $\rho_0$ with a higher precision
 \cite{Catena09, Weber09,
       Salucci10, Pato10, deBoer10}.
 These estimates give rather {\em larger} values for $\rho_0$;
 e.g., Catena and Ullio gave \cite{Catena09}
\beq
   \rho_0
 = 0.39 \pm 0.03~{\rm GeV/cm^3}
\~,
\eeq
 and Salucci {\it et al.} even gave \cite{Salucci10}
\beq
   \rho_0
 = 0.43 \pm 0.11 \pm 0.10~{\rm GeV/cm^3}
\~.
\eeq
 Moreover,
 instead of a spherical symmetric density profile
 assumed in Refs.~\cite{Catena09, Salucci10},
 in Refs.~\cite{Weber09, Pato10, deBoer10}
 the authors considered an axisymmetric density profile
 for a flattened Galactic Dark Matter halo \cite{Sackett94}
 caused by the disk structure of the luminous baryonic component.
 It was found that
 the local density of such a non--spherical Dark Matter halo
 could be enhanced by $\sim$ 20\% or larger
 \cite{Weber09, Pato10}
 and Pato {\it et al.} gave therefore \cite{Pato10}
\beq
   \rho_0
 = 0.466 \pm 0.033 ({\rm stat}) \pm 0.077 ({\rm syst})~{\rm GeV/cm^3}
\~.
\eeq
}:
\beq
   \rho_0
 = 0.2 - 0.8~{\rm GeV / cm^3}
\~.
\label{eqn:rho0_range}
\eeq

 On the other hand,
 in most theoretical models,
 the spin--independent WIMP interaction on nucleus
 with an atomic mass number $A~\gsim~30$ dominates over
 the spin--dependent (SD) interaction \cite{SUSYDM96, Bertone05}.
 Additionally,
 for the lightest supersymmetric neutralino,
 which is perhaps the best motivated WIMP candidate
 \cite{SUSYDM96, Bertone05, Bergstrom09},
 and for all WIMPs which interact primarily through Higgs exchange,
 the SI scalar coupling is approximately the same
 on both protons p and neutrons n \cite{Cotta09}.
 The ``pointlike'' cross section $\sigma_0$ in Eq.~(\ref{eqn:calA})
 can thus be written as
\beqn
           \sigmaSI
 \=        \afrac{4}{\pi} \mrN^2
           \bBig{Z f_{\rm p} + (A - Z) f_{\rm n}}^2
           \non\\
 \eqnsimeq \afrac{4}{\pi} \mrN^2 A^2 |f_{\rm p}|^2
           \non\\
 \=        A^2 \afrac{\mrN}{\mrp}^2 \sigmapSI
\~,
\label{eqn:sigma0SI}
\eeqn
 and the SI WIMP cross section on protons (nucleons)
 can be given as
\beq
   \sigmapSI
 = \afrac{4}{\pi} \mrp^2 |f_{\rm p}|^2
\~,
\label{eqn:sigmapSI}
\eeq
 where $f_{\rm p(n)}$ are the effective
 $\chi \chi {\rm p p (nn)}$ four--point couplings,
 $A$ is the atomic mass number of the target nucleus,
 and $\mrp$ is the reduced mass of
 the WIMP mass $\mchi$ and the proton mass $m_{\rm p}$.
 Here the tiny mass difference between a proton and a neutron
 has been neglected.

 As mentioned in the introduction,
 in our earlier work
 it has been found that
 one could in principle determine $\mchi$
 from direct detection experiments
 with neither a prior knowledge of $\sigma_0$
 nor that of $\rho_0$
 \cite{DMDDmchi-SUSY07, DMDDmchi}.
 Conversely,
 I will show in this article that
 one could also estimate or at least constrain
 the WIMP--nucleon cross section
 from experimental data directly {\em without} knowing $\mchi$,
 but for this estimation
 an assumption about $\rho_0$ is needed.
\subsection{Expression for estimating the SI WIMP--nucleon coupling}
 Our analysis starts from the expression (\ref{eqn:dRdQ})
 for the event rate for the elastic WIMP--nucleus scattering directly.
 By using a time--averaged recoil spectrum,
 and assuming that no directional information exists,
 the normalized one--dimensional
 velocity distribution function of halo WIMPs, $f_1(v)$,
 has been solved from Eq.~(\ref{eqn:dRdQ}) analytically \cite{DMDDf1v}
 and,
 consequently,
 its generalized moments can be estimated by
 \cite{DMDDf1v, DMDDmchi}%
\footnote{
 Here we have implicitly assumed that
 $\Qmax$ is so large that
 terms involving $- 2 \Qmax^{(n+1)/2} r(\Qmax) / F^2(\Qmax)$
 are negligible.
 Due to sizable contributions
 from large recoil energies \cite{DMDDf1v},
 this is not necessarily true,
 especially for some not--very--high $\Qmax$
 in the experimental reality,
 and/or heavy detector targets,
 and/or heavy WIMPs.
 Nevertheless,
 considering the large statistical uncertainties
 due to (very) few events in the highest energy ranges,
 this should practically be a good approximation.
}
\beqn
    \expv{v^n}(v(\Qmin), v(\Qmax))
 \= \int_{v(\Qmin)}^{v(\Qmax)} v^n f_1(v) \~ dv
    \non\\
 \= \alpha^n
    \bfrac{2 \Qmin^{(n+1)/2} r(\Qmin) / \FQmin + (n+1) I_n(\Qmin, \Qmax)}
          {2 \Qmin^{   1 /2} r(\Qmin) / \FQmin +       I_0(\Qmin, \Qmax)}
\~.
\label{eqn:moments}
\eeqn
 Here $v(Q) = \alpha \sqrt{Q}$,
 $Q_{\rm (min, max)}$ are
 the experimental minimal and maximal
 cut--off energies of the data set,
 respectively,
\beq
        r(\Qmin)
 \equiv \adRdQ_{{\rm expt}, \~Q = \Qmin}
\label{eqn:rmin}
\eeq
 is an estimated value of
 the {\em measured} recoil spectrum $(dR/dQ)_{\rm expt}$
 ({\em before} normalized by an experimental exposure, $\cal E$)
 at $Q = \Qmin$,
 and $I_n(\Qmin, \Qmax)$ can be estimated through the sum:
\beq
   I_n(\Qmin, \Qmax)
 = \sum_{a = 1}^{N_{\rm tot}} \frac{Q_a^{(n-1)/2}}{F^2(Q_a)}
\~,
\label{eqn:In_sum}
\eeq
 where the sum runs over all events in the data set
 that satisfy $Q_a \in [\Qmin, \Qmax]$
 and $N_{\rm tot}$ is the number of such events.
 Note that,
 firstly,
 by using the second line of Eq.~(\ref{eqn:moments})
 $\expv{v^n}(v(\Qmin), v(\Qmax))$ can be determined
 independently of the local WIMP density $\rho_0$,
 of the velocity distribution function of incident WIMPs, $f_1(v)$,
 as well as of the WIMP--nucleus cross section $\sigma_0$.
 Secondly,
 $r(\Qmin)$ and $I_n(\Qmin, \Qmax)$
 are two key quantities for our analysis,
 which can be estimated
 either from a functional form of the recoil spectrum
 or from experimental data (i.e., the measured recoil energies) directly%
\footnote{
 All formulae needed for estimating
 $r(\Qmin)$, $I_n(\Qmin, \Qmax)$, and their statistical errors
 are given in the appendix.
%
}.

 By substituting the second expression in Eq.~(\ref{eqn:sigma0SI})
 into Eq.~(\ref{eqn:dRdQ}),
 and using the fact that
 the integral over the one--dimensional WIMP velocity distribution
 on the right--hand side of Eq.~(\ref{eqn:dRdQ})
 is the minus--first moment of this distribution,
 which can be estimated by Eq.~(\ref{eqn:moments}) with $n = -1$,
 we have
\beqn
    \adRdQ_{{\rm expt}, \~Q = \Qmin}
 \= \calE \calA \FQmin \int_{v(\Qmin)}^{v(\Qmax)} \bfrac{f_1(v)}{v} dv
    \non\\
 \= \calE \afrac{2 \rho_0 A^2 |f_{\rm p}|^2}{\pi \mchi} \FQmin \cdot
    \frac{1}{\alpha}
    \bfrac{2 r(\Qmin) / \FQmin}{2 \Qmin^{1/2} r(\Qmin) / \FQmin + I_0}
\~.
\eeqn
 Using the definition (\ref{eqn:alpha}) of $\alpha$,
 the {\em squared} SI WIMP coupling on protons (nucleons)
 can be expressed as
 \cite{DMDDidentification-DMDE2009}
\beq
   |f_{\rm p}|^2
 = \frac{1}{\rho_0}
   \bbrac{\frac{\pi}{4 \sqrt{2}} \afrac{1}{\calE A^2 \sqrt{\mN}}}
   \bbrac{\frac{2 \Qmin^{1/2} r(\Qmin)}{\FQmin} + I_0}
   \abrac{\mchi + \mN}
\~.
\label{eqn:fp2}
\eeq
 Note that,
 firstly,
 the experimental exposure $\calE$
 appearing in the denominator
 relates the {\em actual} counting rate $(dR / dQ)_{\rm expt}$
 to the normalized rate in Eq.~(\ref{eqn:dRdQ}).
 Secondly,
 due to the neglect of the terms
 $- 2 \Qmax^{1/2} r(\Qmax) / F^2(\Qmax)$
 and $- 2 r(\Qmax) / F^2(\Qmax)$
 in the denominator and numerator of
 the expression (\ref{eqn:moments}) for $\expv{v^n}$,
 respectively,
 $|f_{\rm p}|^2$ determined by Eq.~(\ref{eqn:fp2})
 would be {\em overestimated},
 since the contributions from the two neglected terms
 are {\em negative} and
 the former is much larger then the later.
 However,
 because $|f_{\rm p}|^2$ estimated by Eq.~(\ref{eqn:fp2})
 is inversely proportional to the local WIMP density,
 whose commonly used value
 would possibly be {\em underestimated}
 (see Eqs.~(\ref{eqn:rho0}) to (\ref{eqn:rho0_range})),
 one should therefore at least be able to
 give an {\em upper} bound on $|f_{\rm p}|^2$.
 Then,
 by using the standard Gaussian error propagation,
 the statistical uncertainty on $|f_{\rm p}|^2$
 estimated by Eq.~(\ref{eqn:fp2}) can be given as
\beq
    \sigma(|f_{\rm p}|^2)
 = |f_{\rm p}|^2
    \bbrac{  \frac{\sigma^2(\mchi)}{(\mchi + \mN)^2}
           + \calN_{\rm m}^2 \sigma^2(1 / \calN_{\rm m})
           + \frac{2 \calN_{\rm m} \~ {\rm cov}(\mchi, 1 / \calN_{\rm m})}
                  {(\mchi + \mN)}}^{1/2}
\~,
\label{eqn:sigma_fp2}
\eeq
 where I have used \cite{DMDDf1v}
\beq
   \calN_{\rm m}^{-1}
 = \frac{2 \Qmin^{1/2} r(\Qmin)}{\FQmin} + I_0
\~.
\label{eqn:calNm}
\eeq
\subsection{From a single experiment}
 The expression (\ref{eqn:fp2}) for estimating
 the (squared) SI WIMP--proton coupling
 depends on three quantities:
 $r(\Qmin)$, $I_0$, and the WIMP mass $\mchi$.
 As argued in Ref.~\cite{DMDDmchi},
 from a {\em single} recoil spectrum
 one {\em cannot} estimate $\mchi$ 
 {\em without} making some assumptions
 about the velocity distribution $f_1(v)$.
 Hence,
 as a model--independent analysis,
 one could only express/constrain
 $|f_{\rm p}|^2$
 as a (linear) function/interval of the WIMP mass
 on the coupling--mass plane
 by using Eq.~(\ref{eqn:fp2}) with a {\em single} experiment.
 Meanwhile,
 from Eqs.~(\ref{eqn:dRdQ}) and (\ref{eqn:calA}),
 it can be found that,
 due to the degeneracy between
 the local WIMP density $\rho_0$ and
 the WIMP--nucleus cross section $\sigma_0$,
 one {\em cannot} estimate both of them independently%
\footnote{
 In contrast,
 as I will show in Sec.~4,
 the ratios between different WIMP--nucleon couplings/cross sections
 can be determined {\em without} knowing
 the mass and the local WIMP density
 \cite{DMDDidentification-DARK2009,
       DMDDidentification-DMDE2009,
       DMDDranap}.
}.
 Thus,
 for using Eq.~(\ref{eqn:fp2}),
 the simplest way is making
 an assumption for the local WIMP density $\rho_0$.

\begin{figure}[t!]
\begin{center}
\includegraphics[width=15cm]{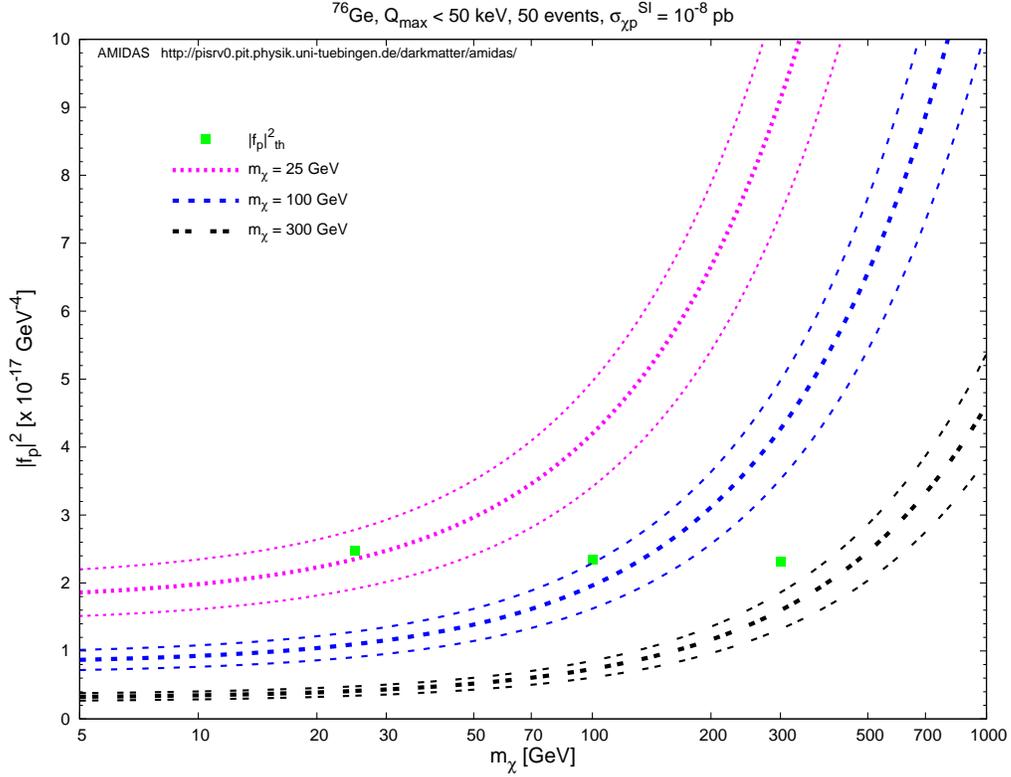} \\
\vspace{-0.25cm}
\end{center}
\caption{
 The {\em squared} SI WIMP--proton couplings $|f_{\rm p}|^2$
 estimated by Eq.~(\ref{eqn:fp2})
 and the lower and upper bounds of
 their 1$\sigma$ statistical uncertainties
 as functions of the WIMP mass
 for a $\rmXA{Ge}{76}$ target.
 The theoretical predicted recoil spectrum for
 the shifted Maxwellian velocity distribution with
 \mbox{$v_0 = 220$ km/s},
 $v_{\rm e} = 1.05 \~ v_0$,
 and \mbox{$\vmax = 700$ km/s}
 as well as the commonly used elastic nuclear form factor
 for the SI cross section
 given in Eq.~(\ref{eqn:FQ_WS})
 have been used.
 The SI WIMP--proton cross section
 has been set as $10^{-8}$ pb.
 The experimental maximal cut--off energy $\Qmax$
 has been set as 50 keV
 and the threshold energy has been assumed to be negligible.
 Each experiment contains 50 total events on average.
 The mass of incident WIMPs has been chosen
 as 25 (dotted magenta),
 100 (dashed blue),
 and 300 (double--dashed black) GeV,
 respectively.
 The filled green squares indicate
 the input WIMP masses and
 the theoretical values of $|f_{\rm p}|^2$.
 See the text for further details.
}
\label{fig:fp2_single_50}
\end{figure}

 In Fig.~\ref{fig:fp2_single_50}
 I show the simulated results
 for a $\rmXA{Ge}{76}$ target
 with 5,000 experiments
 based on the Monte Carlo method%
\footnote{
 Note that,
 rather than the mean values,
 in this article
 we give always the median values
 of the reconstructed results
 from the simulated experiments.
}.
 The theoretical predicted recoil spectrum for
 the shifted Maxwellian velocity distribution
 \cite{SUSYDM96, Bertone05, DMDDf1v} with
 a Sun's orbital velocity in the Galactic frame $v_0 = 220$ km/s,
 an Earth's velocity in the Galactic frame $v_{\rm e} = 1.05 \~ v_0$,%
\footnote{
 The time dependence of the Earth's velocity
 in the Galactic frame \cite{SUSYDM96, Bertone05}
 has been ignored.
}
 and a maximal cut--off velocity of
 the velocity distribution function $\vmax = 700$ km/s,
 as well as the commonly used elastic nuclear form factor
 for the SI cross section
 \cite{Engel91, SUSYDM96, Bertone05}:
\beq
   F_{\rm SI}^2(Q)
 = \bfrac{3 j_1(q R_1)}{q R_1}^2 e^{-(q s)^2}
\label{eqn:FQ_WS}
\eeq
 have been used.
 The SI WIMP--proton cross section
 has been set as $10^{-8}$ pb.
 The commonly used value of $\rho_0 = 0.3~{\rm GeV/cm^3}$
 has been used for both predicting the recoil spectrum
 and analyzing generated events.
 The experimental maximal cut--off energy $\Qmax$
 has been set as \mbox{50 keV}
 and the threshold energy has been assumed to be negligible.
 Each experiment contains an expected number of 50 total events;
 the actual event number is Poisson--distributed
 around this expectation value.
 The mass of incident WIMPs has been chosen
 as 25 (dotted magenta),
 100 (dashed blue),
 and 300 (double--dashed black) GeV,
 respectively.

\begin{figure}[t!]
\begin{center}
\includegraphics[width=15cm]{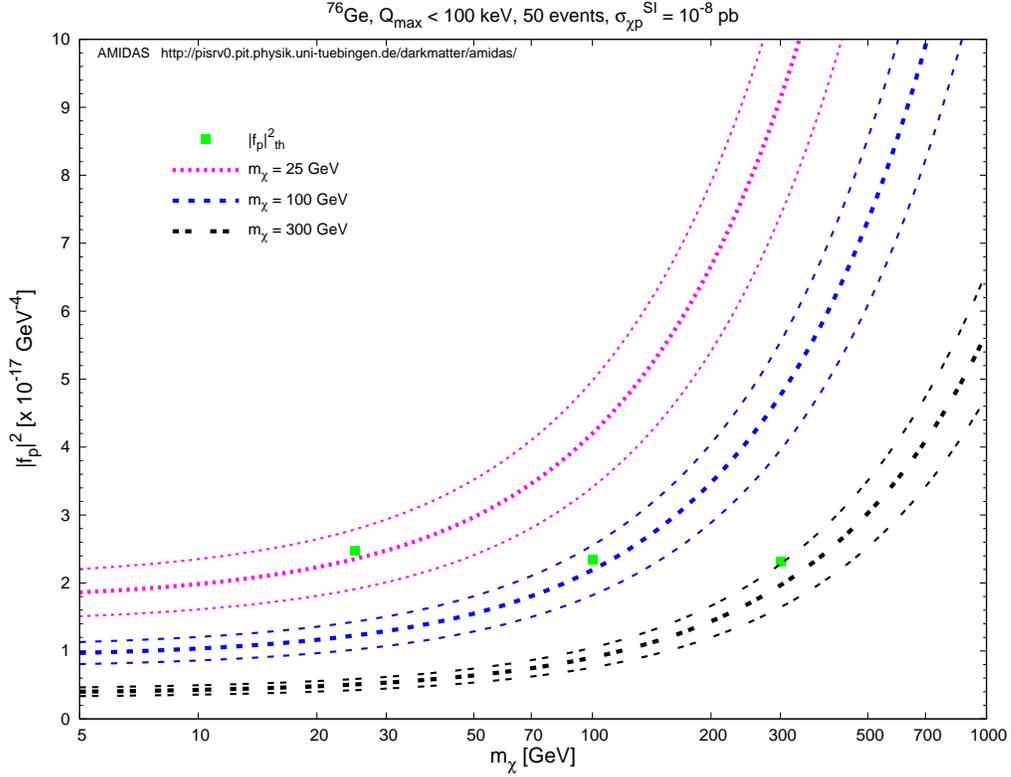} \\
\vspace{-0.6cm}
\end{center}
\caption{
 As in Fig.~\ref{fig:fp2_single_50},
 except that
 the maximal cut--off energy $\Qmax$ has been increased to \mbox{100 keV}.
}
\label{fig:fp2_single}
\end{figure}

 As we can see here,
 the prefactor, i.e., the slope of the linear function
 $|f_{\rm p}|^2(\mchi)$, in Eq.~(\ref{eqn:fp2})
 is obviously {\em underestimated}.
 For the case of an input WIMP mass of 300 GeV,
 the theoretical value of $|f_{\rm p}|^2$
 (the filled green square)
 is even outside the 1$\sigma$ statistical uncertainty interval.
 This is because that
 the experimental maximal cut--off energy
 has been set as only 50 keV here.
 Remind that
 it is usually assumed that
 the WIMP flux on the Earth is negligible
 at velocities exceeding the maximal velocity $\vmax$.
 This leads thus to a kinematic maximum of the recoil energy
\beq
   Q_{\rm max, kin}
 = \frac {\vmax^2} {\alpha^2}
\~.
\label{eqn:Qmax_kin}
\eeq
 For a WIMP mass of 100 (300) GeV,
 this kinematic maximum for a Ge target is 264 (504) keV.
 Hence,
 $I_0$ in the prefactor of the linear function $|f_{\rm p}|^2(\mchi)$
 given in Eq.~(\ref{eqn:fp2})
 has been (strongly) underestimated.
 In Fig.~\ref{fig:fp2_single}
 we increase therefore
 the maximal cut--off energy $\Qmax$ to 100 keV.
 It can be seen clearly that,
 by extending the detector sensitivity
 to higher energy ranges,
 the underestimated $I_0$ and thereby
 the prefactor of the linear function $|f_{\rm p}|^2(\mchi)$
 can be corrected significantly%
\footnote{
 Remind that,
 since we neglected the term $- 2 \Qmax^{1/2} r(\Qmax) / F^2(\Qmax)$
 in the second bracket in Eq.~(\ref{eqn:fp2}),
 which contributes negatively,
 all results shown in this paper
 are somehow {\em overestimated}.
}.

\begin{figure}[p!]
\begin{center}
\vspace{-0.75cm}
\includegraphics[width=15cm]{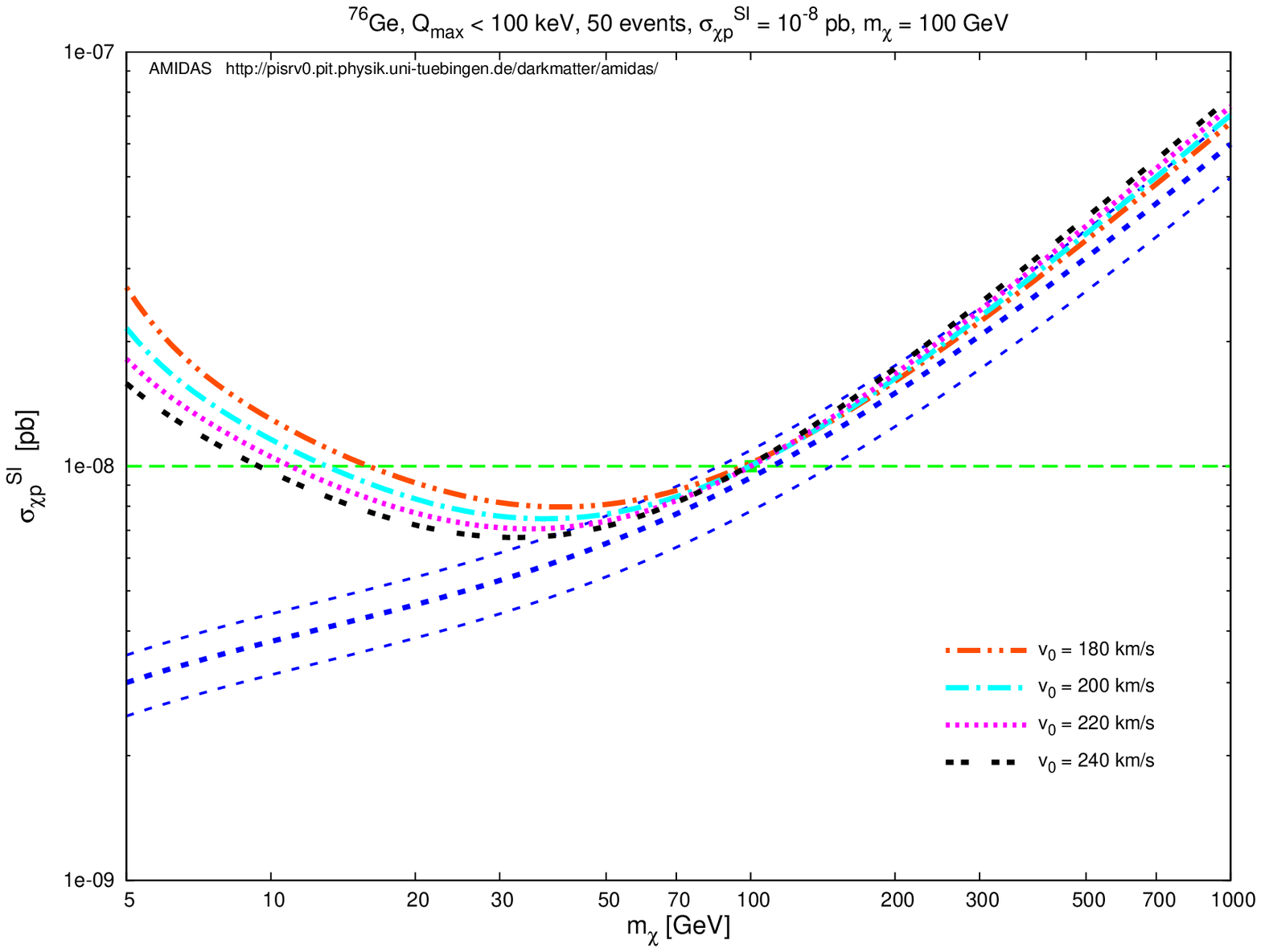}   \\ \vspace{0.25cm}
\includegraphics[width=15cm]{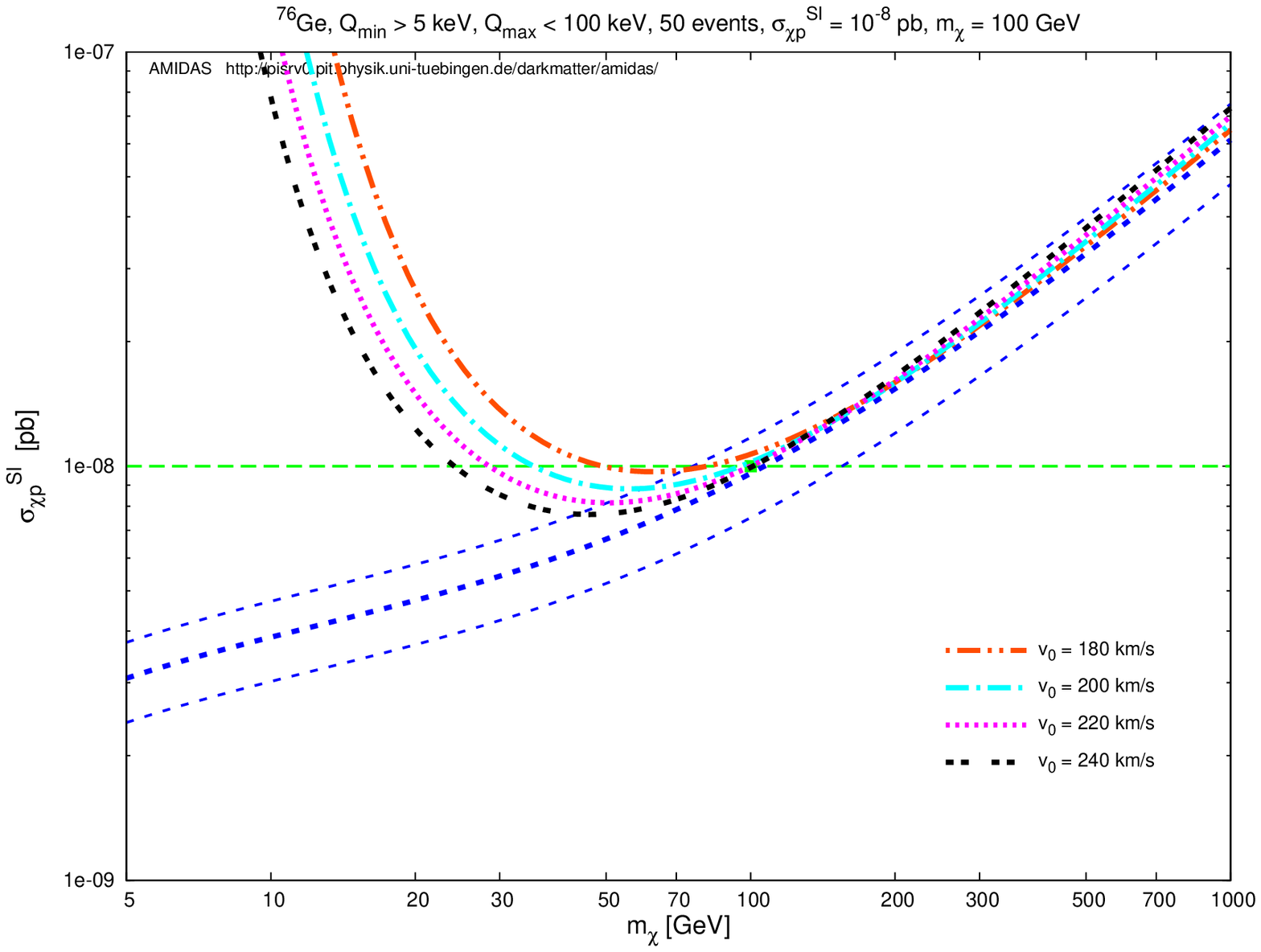} \\
\vspace{-0.5cm}
\end{center}
\caption{
 The SI WIMP--proton cross section $\sigmapSI$
 estimated by Eqs.~(\ref{eqn:fp2}) and (\ref{eqn:sigmapSI})
 and the lower and upper bounds of
 its 1$\sigma$ statistical uncertainty
 as functions of the WIMP mass
 (dashed blue curves)
 for a $\rmXA{Ge}{76}$ target.
 The input WIMP mass is 100 GeV.
 The threshold energies have been set as
 0 (upper) and 5 (lower) keV,
 respectively.
 The four extra curves have been drawn conventionally
 by using the shifted Maxwellian velocity distribution
 with four Sun's orbital velocities:
 \mbox{$v_0 = 180~{\rm km/s}$} (dash--double--dotted orange),
 \mbox{$v_0 = 200~{\rm km/s}$} (dash--dotted cyan),
 \mbox{$v_0 = 220~{\rm km/s}$} (dotted magenta),
 \mbox{$v_0 = 240~{\rm km/s}$} (double--dotted black),
 and the form factor given in Eq.~(\ref{eqn:FQ_WS}).
 The other parameters are as in Fig.~\ref{fig:fp2_single}.
 See the text for further details.
}
\label{fig:sigmaSIp_single}
\end{figure}
\begin{figure}[p!]
\begin{center}
\includegraphics[width=15cm]{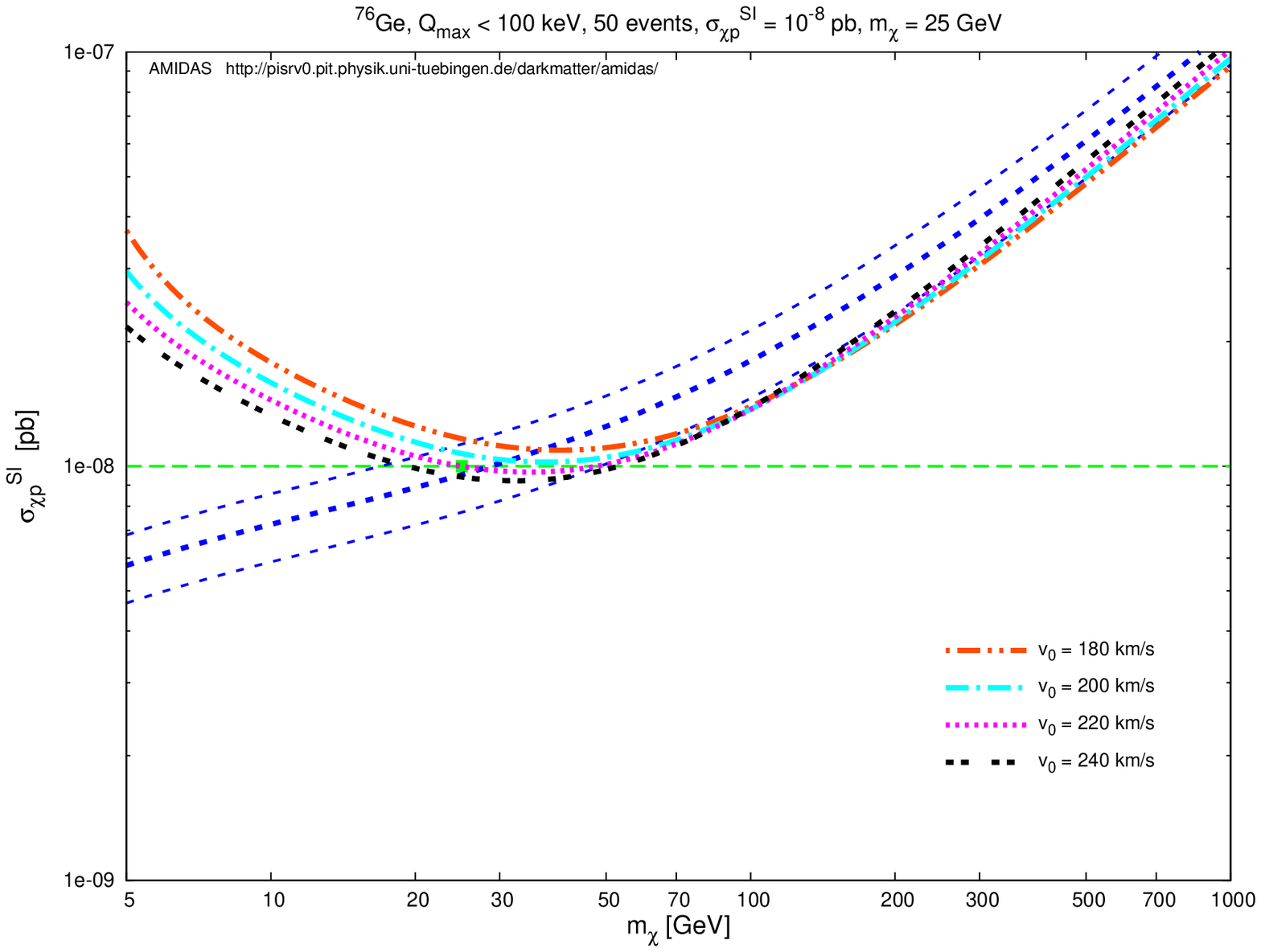}   \\ \vspace{1cm}
\includegraphics[width=15cm]{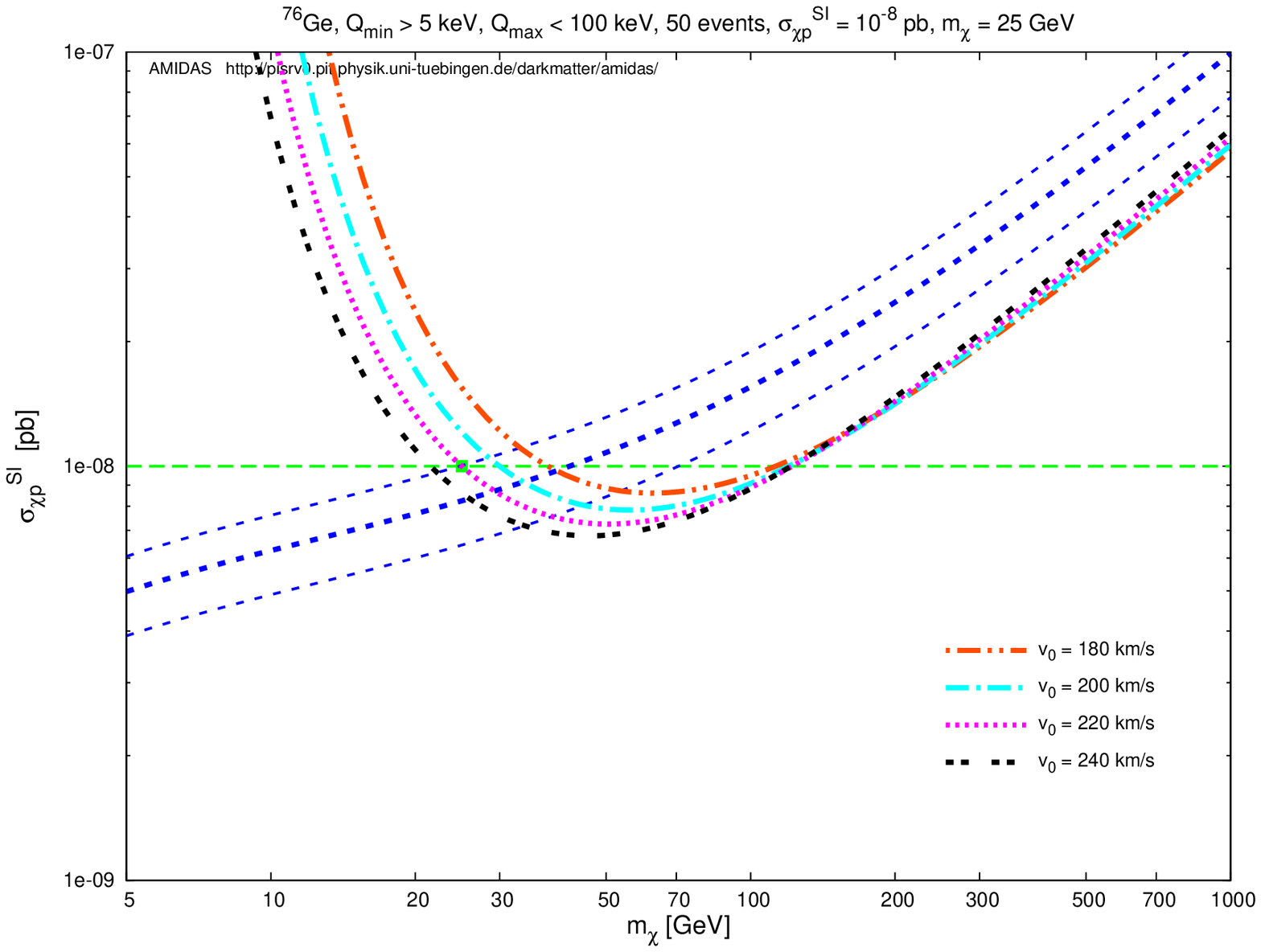} \\
\end{center}
\caption{
 As in Figs.~\ref{fig:sigmaSIp_single},
 except that
 the input WIMP mass is only 25 GeV here.
}
\label{fig:sigmaSIp_single_025}
\end{figure}

 On the other hand,
 by substituting Eq.~(\ref{eqn:fp2}) into Eq.~(\ref{eqn:sigmapSI}),
 one can express the SI WIMP--proton (nucleon) cross section
 as a function of the WIMP mass, $\sigmapSI(\mchi)$,
 on the cross section versus WIMP mass plane.
 In Figs.~\ref{fig:sigmaSIp_single}
 I show the simulated results for a Ge target
 with an input WIMP mass of 100 GeV.
 The experimental minimal cut--off energy $\Qmin$
 has been set as 0 (upper) and 5 (lower) keV.
 As a comparison
 I show also four extra curves drawn conventionally
 by using the shifted Maxwellian velocity distribution
 with four Sun's orbital velocities:
 \mbox{$v_0 = 180~{\rm km/s}$} (dash--double--dotted orange),
 \mbox{$v_0 = 200~{\rm km/s}$} (dash--dotted cyan),
 \mbox{$v_0 = 220~{\rm km/s}$} (dotted magenta),
 \mbox{$v_0 = 240~{\rm km/s}$} (double--dotted black),
 and the form factor given in Eq.~(\ref{eqn:FQ_WS}).

 As shown here,
 two results analyzed by Eq.~(\ref{eqn:fp2})
 and by the conventional method with an assumed halo model
 are compatible with each other
 in the mass and cross section ranges
 around and higher than the input values,
 whereas
 in the low WIMP mass range,
 these two curves show a significant incompatibility.
 Hence,
 by comparing results from these two analyses,
 one could in principle
 -- for the first step with only one experiment
 observing positive signals --
 give the {\em lower} bounds of the WIMP mass
 and its cross section on protons (nucleons)
 (\mbox{$\mchi~\gsim~40$ GeV} and
  \mbox{$\sigmapSI~\gsim~7 \times 10^{-9}$ pb}
  from the upper frame of Figs.~\ref{fig:sigmaSIp_single}
  in our simulation)
 from a {\em single} experiment.
 Moreover,
 the lower frame of Figs.~\ref{fig:sigmaSIp_single}
 shows that,
 due to the {\em non--negligible} threshold energy
 the conventional method is (much) more unsensitive for lighter WIMPs
 (in contrast,
  the uncertainty interval
  given by Eqs.~(\ref{eqn:fp2}) and (\ref{eqn:sigma_fp2})
  becomes only a bit wider)
 and the curves thus go sharply upwards
 as the WIMP mass decreases.
 The incompatibility between two analyses becomes larger
 and one could therefore even
 give more strict constraints
 on the WIMP mass and the SI cross section
 (\mbox{$\mchi~\gsim~45$ GeV} and
  \mbox{$\sigmapSI~\gsim~7.5 \times 10^{-9}$ pb}
  in our simulation).

 In Figs.~\ref{fig:sigmaSIp_single_025}
 we examine the same comparison of two analyses
 for a rather light input WIMP mass of 25 GeV.
 The lower frame shows that,
 with the non--negligible threshold energy
 one could even give the {\em upper} bounds of
 the WIMP mass and its cross section on protons (nucleons)
 (\mbox{20 GeV} $\lsim~\mchi~\lsim$ \mbox{50 GeV} and
  \mbox{$7 \times 10^{-9}$ pb} $\lsim~\sigmapSI~\lsim$ \mbox{$1.1 \times 10^{-8}$ pb}
  in our simulation)
 from a single experiment.
 However,
 remind that
 with a fixed maximal cut--off energy
 and a number of total events,
 the higher the threshold energy,
 the larger the required exposure.
 Moreover,
 as we can see in the lower frame of
 Figs.~\ref{fig:sigmaSIp_single_025},
 the prefactor, or equivalently, $I_0$, in Eq.~(\ref{eqn:fp2})
 is {\em underestimated} due to a low
 (in contrast to the case shown
  in Figs.~\ref{fig:fp2_single_50} and \ref{fig:fp2_single})
 kinematic maximum of the recoil energy.
 For a WIMP mass of 25 GeV,
 this kinematic maximum for a Ge target is 52.6 keV.
 Remind also that the recoil energy spectrum
 is approximately exponential,
 thus between $Q = 0$ and \mbox{$Q = Q_{\rm max, kin} =$ 52.6 keV},
 only $\sim$ 53\% of the total events
 are with energies \mbox{$Q \ge \Qmin = 5$ keV}.
 Due to this underestimate of $I_0$,
 we will see later that
 the non--negligible threshold energy
 could cause serious problem once the WIMPs are (pretty) light.

\begin{figure}[p!]
\begin{center}
\vspace{-0.5cm}
\includegraphics[width=15cm]{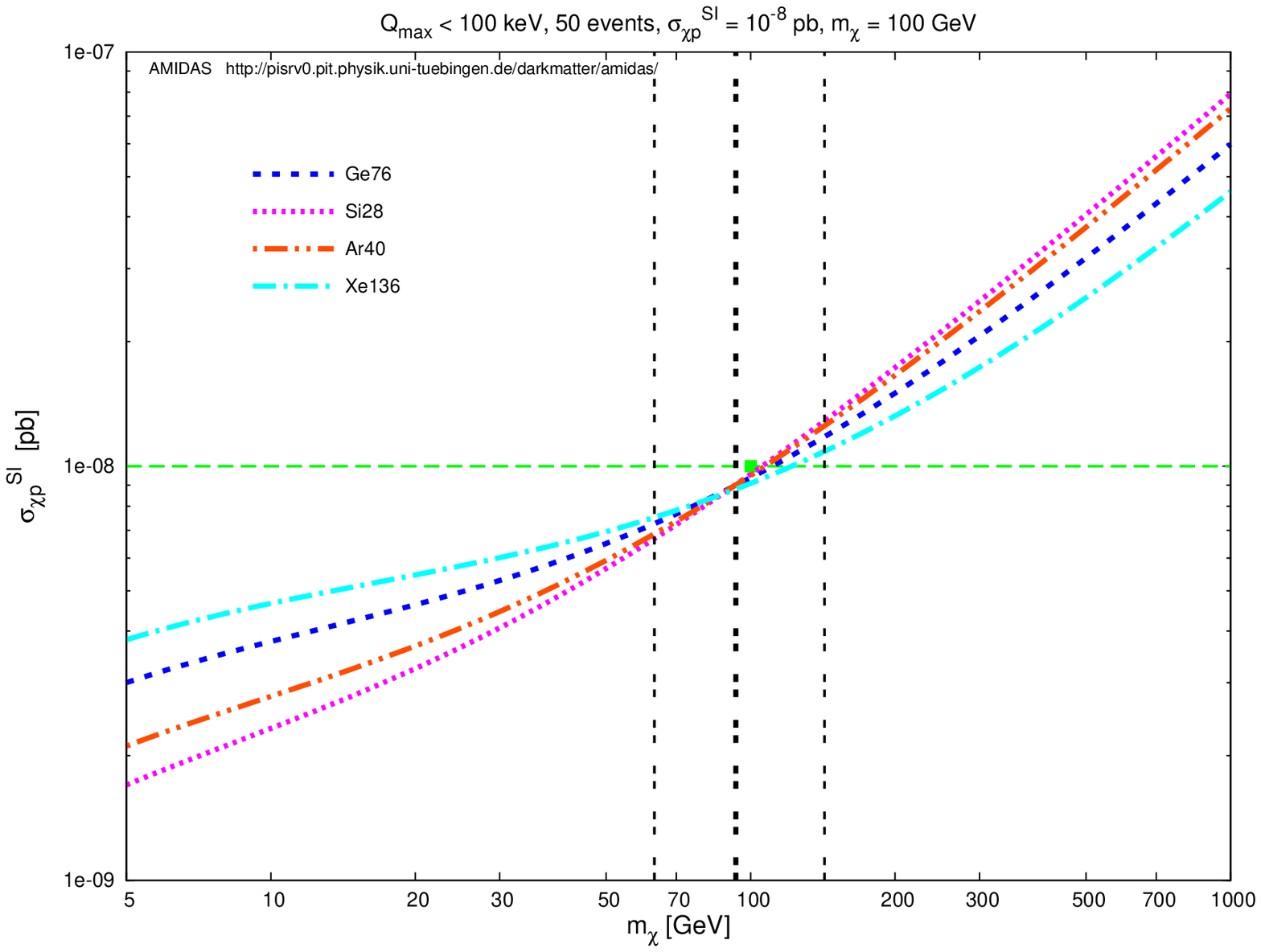}     \\ \vspace{0.25cm}
\includegraphics[width=15cm]{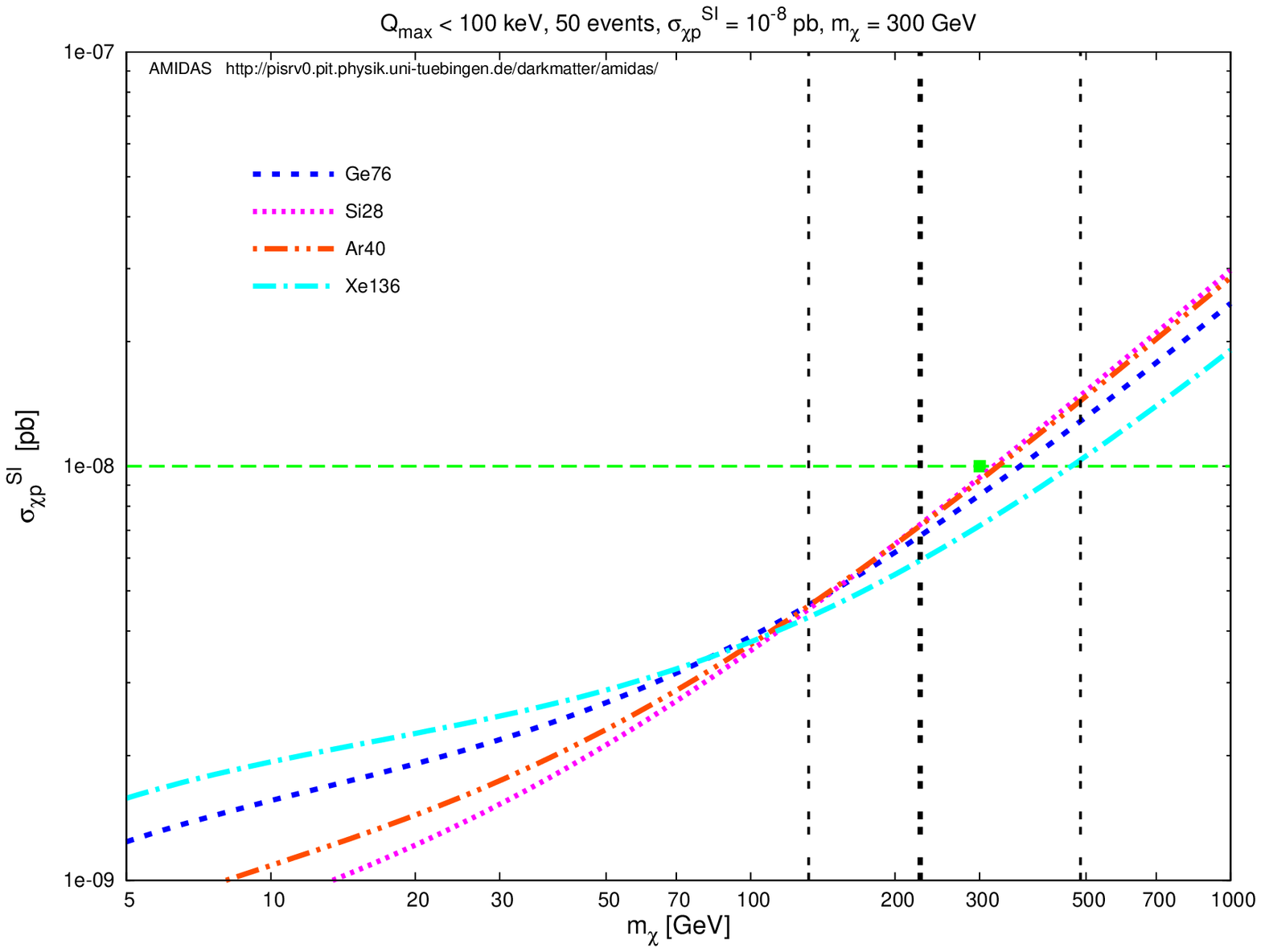} \\
\vspace{-0.5cm}
\end{center}
\caption{
 The SI WIMP--proton cross sections $\sigmapSI(\mchi)$
 estimated by Eqs.~(\ref{eqn:fp2}) and (\ref{eqn:sigmapSI})
 as functions of the WIMP mass
 for four different target nuclei:
 $\rmXA{Ge}{76}$ (dashed blue),
 $\rmXA{Si}{28}$ (dotted magenta),
 $\rmXA{Ar}{40}$ (dash--double--dotted orange),
 and $\rmXA{Xe}{136}$ (long--dash-dotted cyan).
 The input WIMP mass
 has been set as 100 (upper) and 300 (lower) GeV.
 The threshold energies for all targets
 are assumed to be negligible.
 The vertical double--dashed black lines
 show the reconstructed WIMP masses
 and the lower and upper bounds of
 their 1$\sigma$ statistical uncertainties
 estimated by the algorithmic procedure
 introduced in Ref.~\cite{DMDDmchi}.
 The other parameters are as in Fig.~\ref{fig:fp2_single}.
}
\label{fig:sigmaSIp_4T}
\end{figure}
\section{Estimating the SI WIMP--nucleon coupling}
 In this section
 I consider further the case that
 two (or more) experiments with different target nuclei
 observe positive WIMP signals.
\subsection{Combining different experiments}

 In Figs.~\ref{fig:sigmaSIp_4T}
 I show the SI WIMP--proton cross sections $\sigmapSI(\mchi)$
 estimated by Eqs.~(\ref{eqn:fp2}) and (\ref{eqn:sigmapSI})
 as functions of the WIMP mass
 on the $\sigmapSI - \mchi$ plane
 for four different target nuclei:
 $\rmXA{Ge}{76}$ (dashed blue),
 $\rmXA{Si}{28}$ (dotted magenta),
 $\rmXA{Ar}{40}$ (dash--double--dotted orange),
 and $\rmXA{Xe}{136}$ (long--dash-dotted cyan).
 Not surprisingly,
 all four curves pass through (approximately)
 the same values of $\mchi$ and $\sigmapSI$.
 It is in fact one of the basic ideas of
 the model--independent determination of
 the WIMP mass \cite{DMDDmchi-SUSY07, DMDDmchi}
 mentioned in the introduction%
\footnote{
 A brief review of the determination of the WIMP mass
 is given in the appendix.
}.
 However,
 one can also find here that
 the (approximately) common values of $\mchi$ and $\sigmapSI$
 are somehow {\em underestimated},
 especially for the heavier input WIMP mass
 (see the lower frame).
 In Ref.~\cite{DMDDmchi},
 we discussed this phenomenon
 and introduced therefore an algorithmic procedure
 to correct this systematic deviation
 by matching the maximal cut--off energies of different targets.
 In Figs.~\ref{fig:sigmaSIp_4T}
 the vertical double--dashed black lines
 show the reconstructed WIMP masses
 and the lower and upper bounds of
 their 1$\sigma$ statistical uncertainties
 estimated by this algorithmic procedure.

\begin{figure}[p!]
\begin{center}
\hspace*{-1.6cm}
\includegraphics[width=9.8cm]{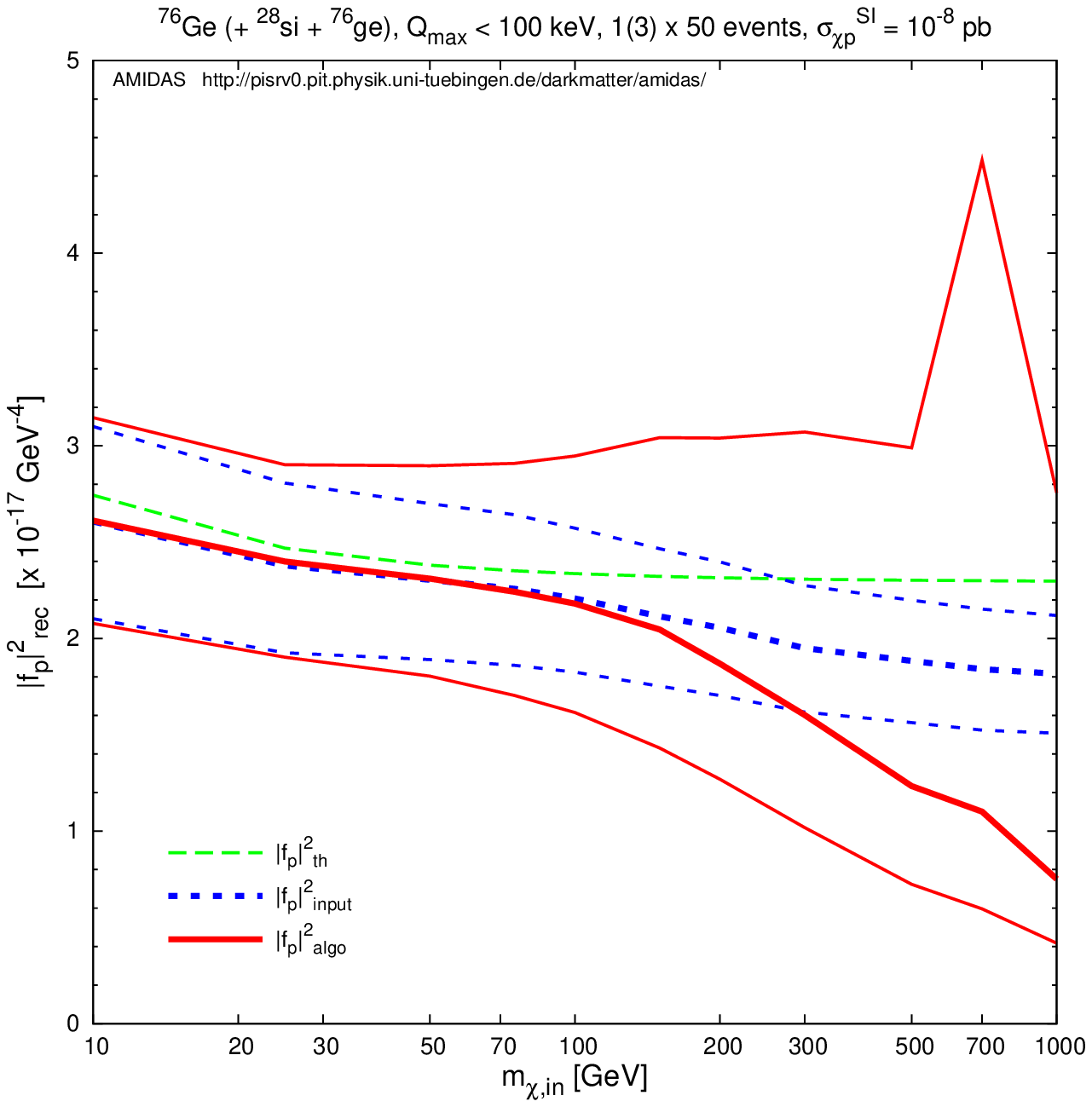} \hspace{-1.1cm}
\includegraphics[width=9.8cm]{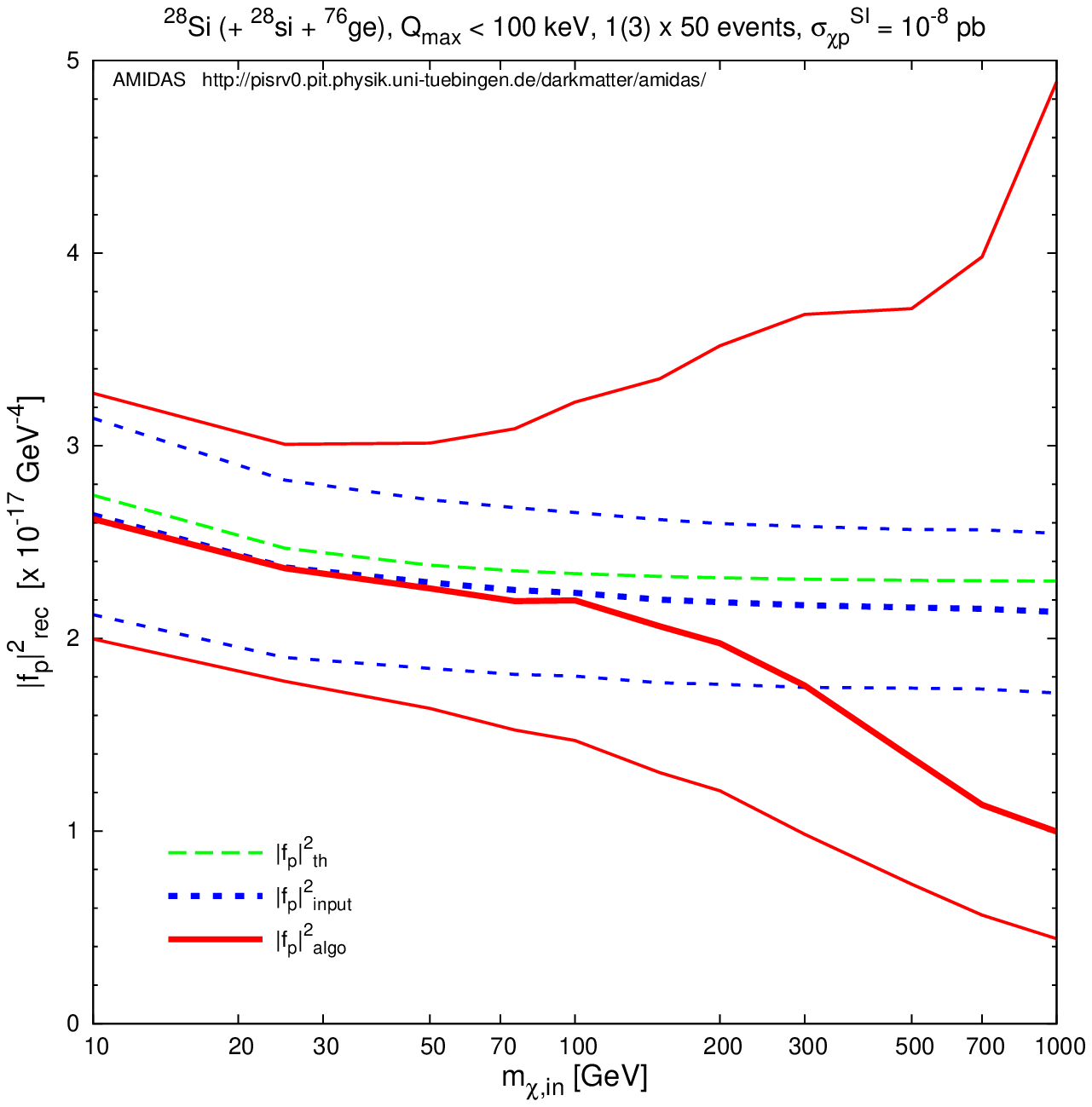} \hspace*{-1.6cm} \\
\vspace{0.75cm}
\hspace*{-1.6cm}
\includegraphics[width=9.8cm]{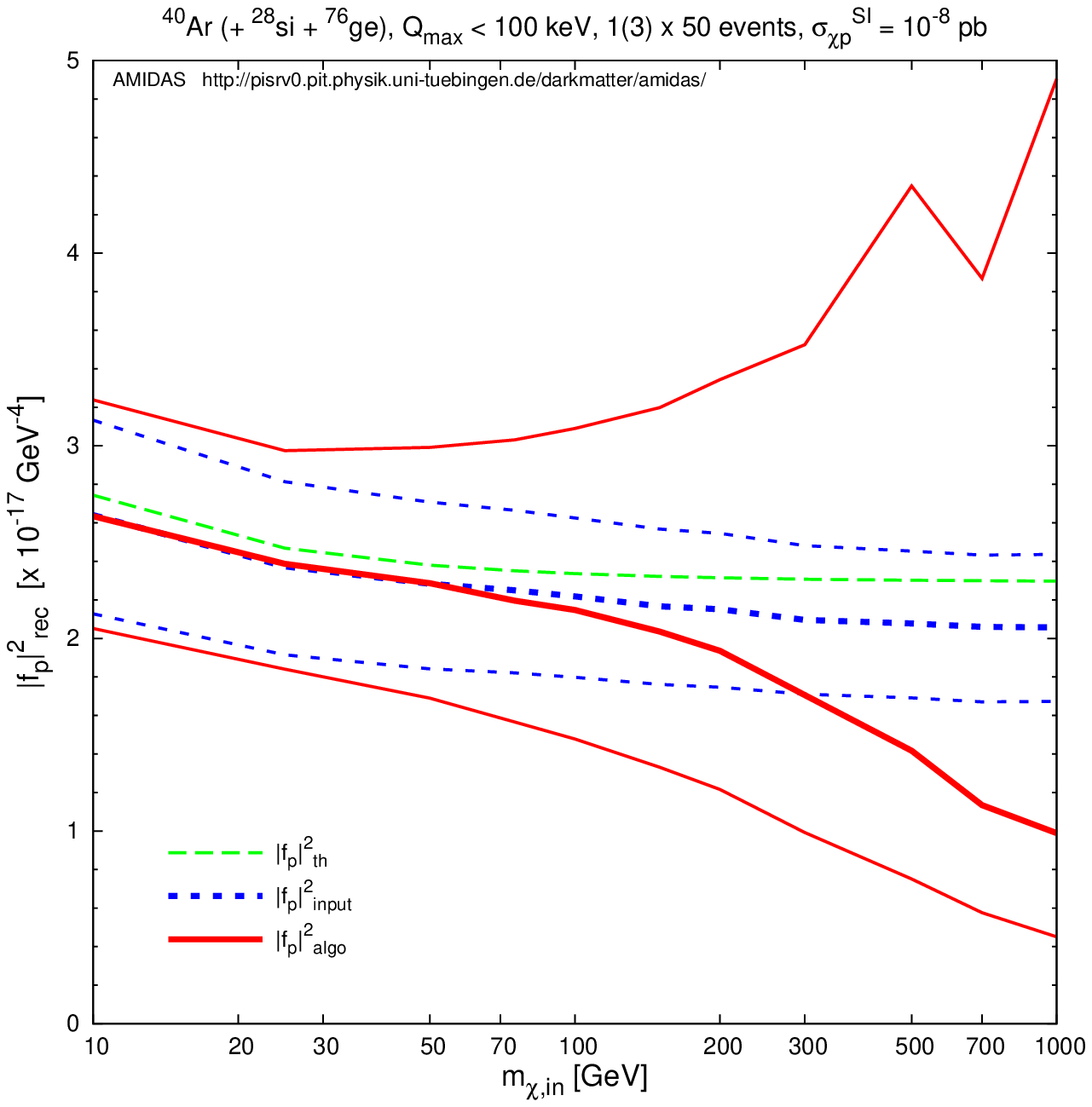} \hspace{-1.1cm}
\includegraphics[width=9.8cm]{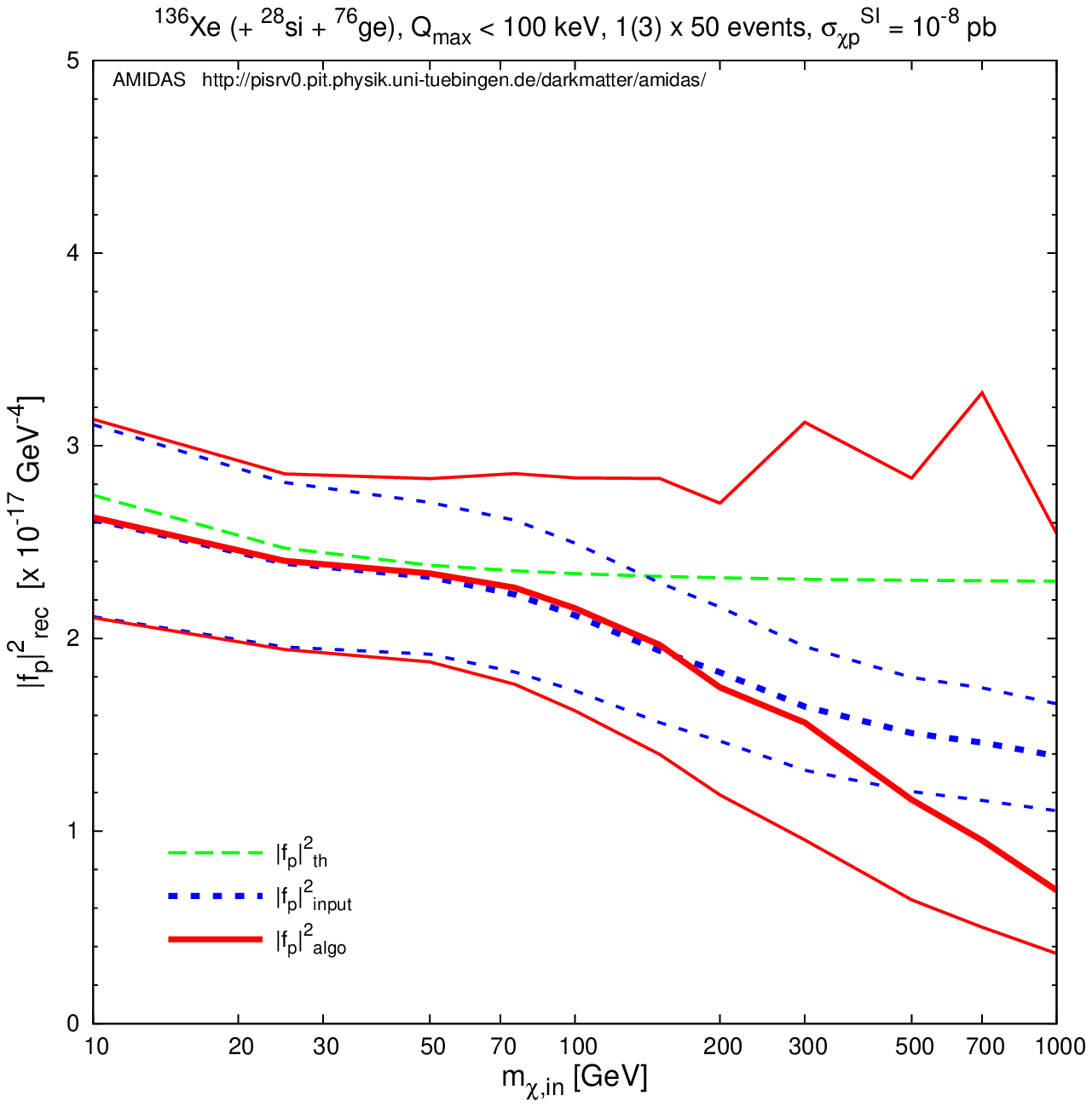} \hspace*{-1.6cm} \\
\end{center}
\caption{
 The reconstructed SI WIMP--proton couplings $|f_{\rm p}|_{\rm rec}^2$
 and the lower and upper bounds of
 their 1$\sigma$ statistical uncertainties
 as functions of the {\em input} WIMP mass $m_{\chi, {\rm in}}$.
 The long--dashed green curve indicates
 the theoretical value of the SI coupling.
 The solid red and dashed blue curves indicate
 the reconstructed SI couplings
 estimated with the reconstructed
 and the input (with an overall uncertainty of 5\%) WIMP masses.
 $\rmXA{Ge}{76}$, $\rmXA{Si}{28}$,
 $\rmXA{Ar}{40}$, and $\rmXA{Xe}{136}$ four nuclei
 have been chosen for estimating
 $r(\Qmin)$ and $I_0$ in Eq.~(\ref{eqn:fp2}).
 Following our work in Ref.~\cite{DMDDmchi},
 $\rmXA{Si}{28}$ and $\rmXA{Ge}{76}$
 (labeled in the plots with small letters
  for their chemical symbols
  to indicate the independence of these data sets
  of the data set of the first nucleus)
 have been chosen as two target nuclei
 for reconstructing the WIMP mass $\mchi$.
 Parameters are as in Figs.~\ref{fig:sigmaSIp_4T}.
 Note that
 for \mbox{$m_{\chi, {\rm in}} \ge$ 500 GeV}
 the upper bounds of the statistical uncertainty
 are systematically {\em underestimated}.
 See the text for further details.
}
\label{fig:fp2_rec}
\end{figure}

 Once the WIMP mass $\mchi$
 on the right--hand side of Eq.~(\ref{eqn:fp2})
 can be determined by means of
 the model--independent method
 with two different target nuclei,
 one can estimate the SI WIMP--proton
 coupling (cross section) straightforwardly.
 Here $r(\Qmin)$ and $I_0$ in the prefactor
 can be estimated from either
 one of the data sets used for determining $\mchi$
 or a third (independent) experiment.
 In Figs.~\ref{fig:fp2_rec},
 I show the reconstructed spin--independent
 WIMP--proton coupling $|f_{\rm p}|_{\rm rec}^2$
 as a function of the {\em input} WIMP mass $m_{\chi, {\rm in}}$.
 The expected number of total events in each data set
 has been set as 50 events on average
 under the experimental maximal cut--off energy $\Qmax$
 set as 100 GeV for all targets;
 the experimental threshold energies
 are assumed to be negligible.
 Four nuclei:
 $\rmXA{Ge}{76}$, $\rmXA{Si}{28}$,
 $\rmXA{Ar}{40}$, and $\rmXA{Xe}{136}$
 have been chosen for estimating
 $r(\Qmin)$ and $I_0$ in Eq.~(\ref{eqn:fp2}).
 Following our work on
 the determination of the WIMP mass \cite{DMDDmchi},
 $\rmXA{Si}{28}$ and $\rmXA{Ge}{76}$
 have been chosen as two target nuclei
 for estimating $\mchi$ in Eq.~(\ref{eqn:fp2}).

\begin{figure}[p!]
\begin{center}
\hspace*{-1.6cm}
\includegraphics[width=9.8cm]{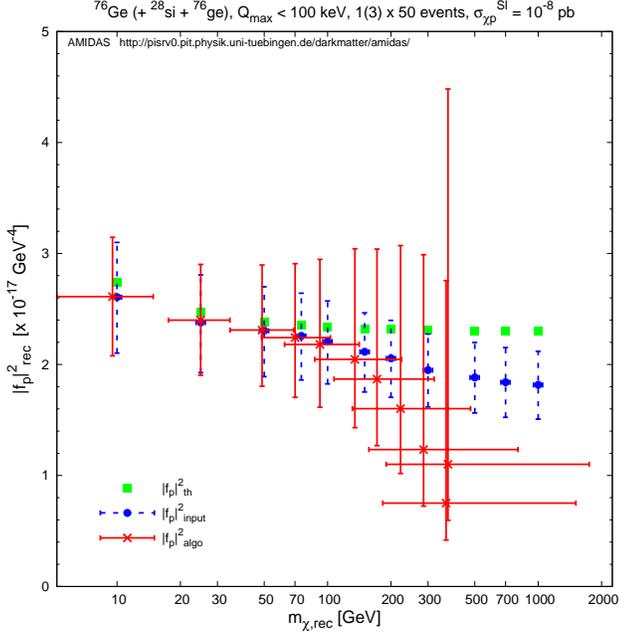} \hspace{-1.1cm}
\includegraphics[width=9.8cm]{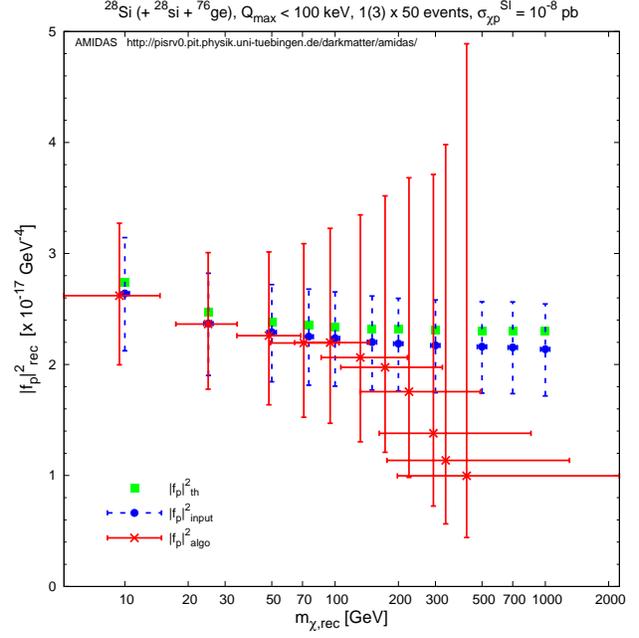} \hspace*{-1.6cm} \\
\vspace{1cm}
\hspace*{-1.6cm}
\includegraphics[width=9.8cm]{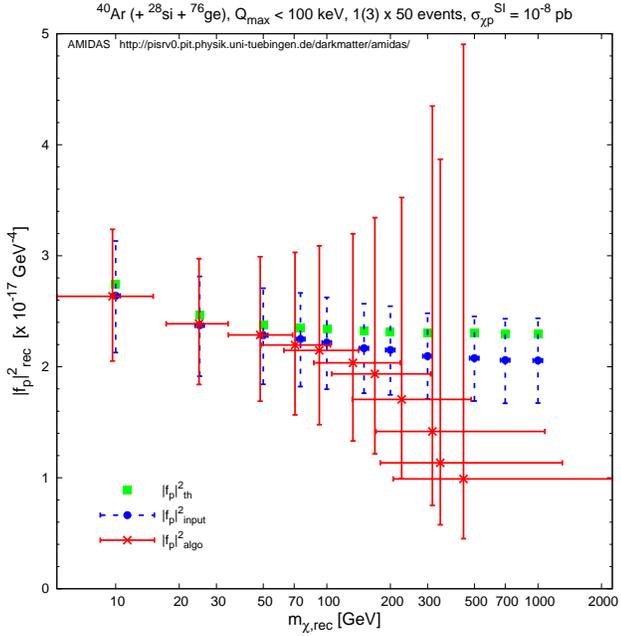} \hspace{-1.1cm}
\includegraphics[width=9.8cm]{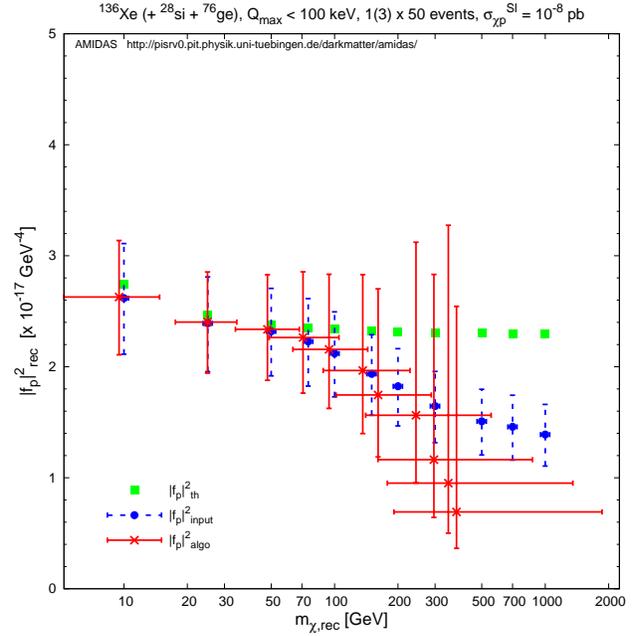} \hspace*{-1.6cm} \\
\end{center}
\caption{
 The reconstructed SI WIMP--proton couplings $|f_{\rm p}|_{\rm rec}^2$
 and the {\em reconstructed} WIMP mass $m_{\chi, {\rm rec}}$
 estimated by the method described in Ref.~\cite{DMDDmchi}
 with the Si and Ge targets
 on the cross section (coupling) versus WIMP mass plane.
 The filled green squares indicate the input WIMP masses and
 the theoretical values of the SI coupling.
 The red crosses (filled blue circles) indicate
 the reconstructed (input) WIMP masses and
 the reconstructed SI couplings
 estimated with these WIMP masses.
 The horizontal (vertical) solid red
 and dashed blue lines
 show the 1$\sigma$ statistical uncertainties
 on $m_{\chi, {\rm rec}}$ ($|f_{\rm p}|_{\rm rec}^2$).
 Parameters are as in Fig.~\ref{fig:fp2_rec}.
}
\label{fig:fp2_mchi_rec}
\end{figure}

 As a comparison to the use of
 the reconstructed WIMP mass (solid red),
 we consider here also the case that
 the WIMP mass $\mchi$ in Eq.~(\ref{eqn:fp2})
 can be determined from some other (collider) experiments
 (dashed blue) with a higher precision.
 The input (true) WIMP mass has been used
 with an overall uncertainty
 of 5\% for this case.
 Note that,
 firstly,
 in order to avoid complicated calculations of the correlations
 between the uncertainty on $\mchi$ estimated by the algorithmic procedure
 and those on $r(\Qmin)$ and $I_0$,
 we have assumed here that
 the two data sets with the Ge (Si) nucleus
 are independent of each other%
\footnote{
 The formulae needed for calculating the correlations
 between the uncertainties
 on the prefactor and on the WIMP mass
 estimated by two basic expressions
 given in Eqs.~(\ref{eqn:mchi_Rn}) and (\ref{eqn:mchi_Rsigma})
 ({\em not} by the algorithmic procedure)
 are given in the appendix.
}.
 Secondly,
 an upper cut--off limit on the reconstructed  WIMP mass
 has been set as \mbox{3000 GeV} in our simulation.
 But,
 due to the very few number of events,
 the {\em upper} bounds of
 the 1$\sigma$ statistical uncertainty
 on the reconstructed mass for heavier input masses
 excess this limit.
 Hence,
 in the heavy mass range \mbox{($m_{\chi, {\rm in}} \ge 500$ GeV)}
 in Figs.~\ref{fig:fp2_rec}
 (and also in Figs.~\ref{fig:fp2_mchi_rec})
 the upper bounds of the 1$\sigma$ statistical uncertainty
 on the reconstructed SI couplings with all four targets
 are {\em systematically underestimated}.

 It can however be found in Figs.~\ref{fig:fp2_rec} that,
 firstly,
 the reconstructed coupling $|f_{\rm p}|_{\rm rec}^2$
 estimated with the input (true) WIMP mass (dashed blue curves)
 for all four targets are {\em underestimated}
 for WIMP masses $\mchi~\gsim~100$ GeV;
 for the heavier target nuclei, Ge and Xe,
 this deviation is larger than for the lighter nuclei, Si and Ar.
 This is caused by the underestimate of $I_0$ in Eq.~(\ref{eqn:fp2}),
 which we found in Figs.~\ref{fig:fp2_single_50} and \ref{fig:fp2_single}
 and discussed there.
 Secondly,
 due to an {\em underestimate} of the reconstructed WIMP mass%
\footnote{
 The WIMP mass has been recostructed by a different program
 than that used in Ref.~\cite{DMDDmchi}.
},
 the reconstructed couplings $|f_{\rm p}|_{\rm rec}^2$
 with the reconstructed WIMP mass (solid red curves)
 for all four targets are more strongly underestimated
 than those with the true WIMP mass
 for WIMP masses $\mchi~\gsim~100$ GeV.
 Moreover,
 for lighter WIMP masses ($m_{\chi, {\rm in}}~\lsim$ 100 GeV),
 the reconstructed $|f_{\rm p}|_{\rm rec}^2$
 for all targets are also a bit {\em underestimated}.
 This could possibly be caused by the statistical fluctuation
 due to the pretty few events.

 The systematic deviation of the reconstructed couplings
 with both reconstructed and real WIMP mass
 caused by the underestimate of $I_0$ reflects the fact that
 the heavier the target nucleus,
 the more the contribution
 from WIMPs with higher velocities
 to the recoil spectrum.
 This observation implies that
 lighter nuclei could be better for estimating $I_0$,
 and in turn for reconstructing the SI WIMP--nucleon coupling.
 This can be seen more clearly
 from the difference of the reconstructed $|f_{\rm p}|^2$
 with Si, Ar and Ge, Xe
 for the case with the input (true) WIMP mass
 (blue dashed curves).
 However,
 Figs.~\ref{fig:fp2_rec} show also that
 the statistical uncertainties on $|f_{\rm p}|^2$
 estimated with the lighter nuclei
 are a bit larger than those with the heavier nuclei.
 And for heavier WIMP masses
 the deviation of $I_0$ could in principle be alleviated
 by extending the experimental maximal cut--off energy $\Qmax$
 to higher energy ranges,
 as discussed in the previous section.
 Moreover,
 remind that
 we simulated here with the same expected event number
 for all four target nuclei.
 In practice,
 we could measure (much) less WIMP events
 in experiments with lighter target nuclei
 ($dR /dQ \propto A^2$).
 This indicates also a larger statistical uncertainty.

 Nevertheless,
 our simulations shown in Figs.~\ref{fig:fp2_rec}
 demonstrate that,
 firstly,
 in spite of the systematic deviation for heavier WIMP masses
 due to the underestimate of $I_0$,
 the true value of $|f_{\rm p}|^2$
 always lies within the 1$\sigma$ statistical uncertainty intervals.
 Secondly,
 for a WIMP mass of \mbox{100 GeV},
 one could in principle estimate
 the squared SI WIMP--proton coupling
 with a statistical uncertainty of $\sim 40\%$ for the Si and Ar targets
 or of only $\sim 30\%$ for the Ge and Xe targets
 with only 50 events from one experiment%
\footnote{
 Note that
 these uncertainties have been estimated by
\beq
 \frac{\vBig{  1\sigma~{\rm upper/lower~bound~of}~|f_{\rm p}|^2
              - |f_{\rm p}|^2_{\rm rec}}}
      {|f_{\rm p}|^2_{\rm rec}}
\~,
\eeq
 and,
 as shown in Figs.~\ref{fig:fp2_rec}
 as well as in Figs.~\ref{fig:fp2_mchi_rec},
 are asymmetric
 since the upper/lower uncertainties
 on the reconstructed WIMP mass
 are asymmetric \cite{DMDDmchi}.
}.
 This is much smaller than the uncertainty on
 the estimate of the local Dark Matter density
 (of a factor of 2 or even larger).

\subsection{Constraints on the cross section--mass plane}

 Since
 by using two (or three) data sets
 the squared SI WIMP--nucleon coupling $|f_{\rm p}|^2$
 can be estimated from experimental data directly
 without knowing the true value of the WIMP mass $\mchi$,
 and the same data sets can also be used
 to reconstruct $\mchi$,
 one can practically
 combine the reconstructed $|f_{\rm p}|^2$
 with the reconstructed $\mchi$ together
 on the cross section--mass plane.
 In Figs.~\ref{fig:fp2_mchi_rec}
 I show the reconstructed SI coupling $|f_{\rm p}|_{\rm rec}^2$
 and the {\em reconstructed} WIMP mass $m_{\chi, {\rm rec}}$
 estimated by the algorithmic procedure
 described in Ref.~\cite{DMDDmchi}
 with the Si and Ge targets
 on the cross section (coupling) versus WIMP mass plane.
 The horizontal (vertical) solid red
 and long--dashed blue lines
 show the 1$\sigma$ statistical uncertainties
 on $m_{\chi, {\rm rec}}$ ($|f_{\rm p}|_{\rm rec}^2$).
 It can be seen that
 the 1$\sigma$ statistical uncertainty areas
 of the reconstructed WIMP mass and its coupling
 can always cover their true values
 up to an input mass of $\sim$ 1 TeV,
 although both of them are underestimated.

 The emphasis here is that,
 while by the conventional analyses
 for determining the WIMP mass and its SI coupling on nucleons
 (see e.g., \cite{Green-mchi, Bernal08, DMDDmchi-NJP})
 one needs a model of the velocity distribution of halo WIMPs,
 one can estimate $\mchi$ and $|f_{\rm p}|^2$ {\em separately}
 by the method presented here
 with {\em neither} prior knowledge of each other
 {\em nor} an assumption about
 the WIMP velocity distribution.
 Certainly,
 how well one can estimate these two quantities
 depends not only on the event number
 but also on the target nucleus,
 as discussed
 in Ref.~\cite{DMDDmchi}
 and shown in Figs.~\ref{fig:fp2_rec} and \ref{fig:fp2_mchi_rec}.

\begin{figure}[p!]
\begin{center}
\hspace*{-1.6cm}
\includegraphics[width=9.8cm]{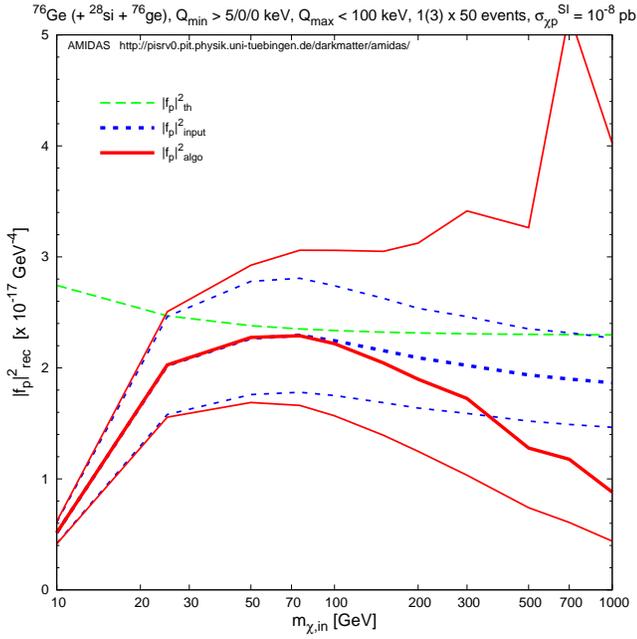}      \hspace{-1.1cm}
\includegraphics[width=9.8cm]{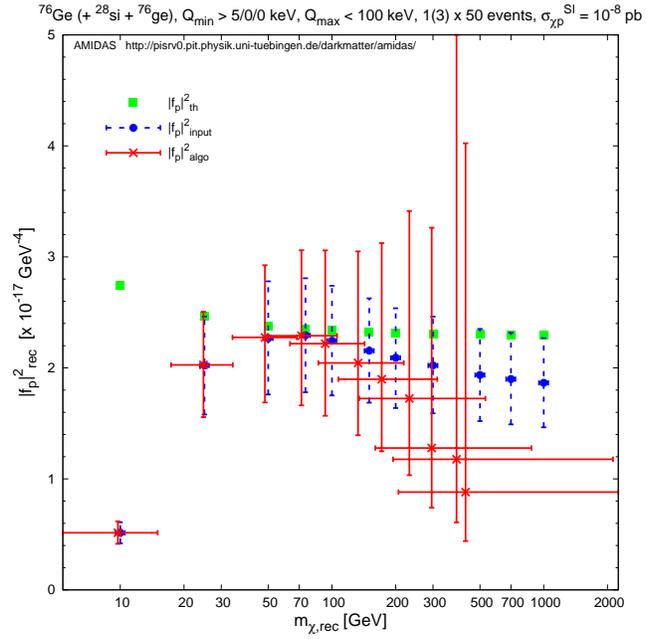} \hspace*{-1.6cm}
\vspace{-0.25cm}
\end{center}
\caption{
 As in Figs.~\ref{fig:fp2_rec} and \ref{fig:fp2_mchi_rec},
 except that
 the experimental minimal cut--off energy
 for the {\em first Ge} target
 has been set as 5 keV.
}
\label{fig:fp2_rec_Qthre}
\end{figure}

 In Figs.~\ref{fig:sigmaSIp_single} and
 \ref{fig:sigmaSIp_single_025}
 we saw that
 the non--negligible threshold energy
 could allow us to give more strict constraints
 on the WIMP mass and its SI coupling on nucleons.
 In Figs.~\ref{fig:fp2_rec_Qthre}
 we therefore take into account
 a minimal cut--off energy $\Qmin = 5$ keV
 for the {\em first Ge} target
 used for estimating $r(\Qmin)$ and $I_0$.
 It can be seen obviously that,
 for lighter WIMP masses ($\mchi \lsim$ 50 GeV)
 the reconstructed coupling $|f_{\rm p}|_{\rm rec}^2$
 is (strongly) {\em underestimated}
 for both cases with the reconstructed
 and the input (true) WIMP masses.
 As discussed at the end of the previous section,
 this is caused by a (very) low
 kinematic maximum of the recoil energy and,
 consequently,
 the underestimate of $I_0$.
 For a WIMP mass of 10 GeV,
 this kinematic maximum is just 11.8 keV and
 between $Q = 0$ and \mbox{$Q = Q_{\rm max, kin} =$ 11.8 keV},
 only $\sim$ 6.4\% of the total events
 are with energies \mbox{$Q \ge \Qmin = 5$ keV}!
 In contrast,
 for heavier WIMP masses ($\mchi~\gsim$ 50 GeV),
 the non--negligible threshold energy
 causes only slightly larger statistical uncertainties
 on the reconstructed SI couplings.

\begin{figure}[p!]
\begin{center}
\hspace*{-1.6cm}
\includegraphics[width=9.8cm]{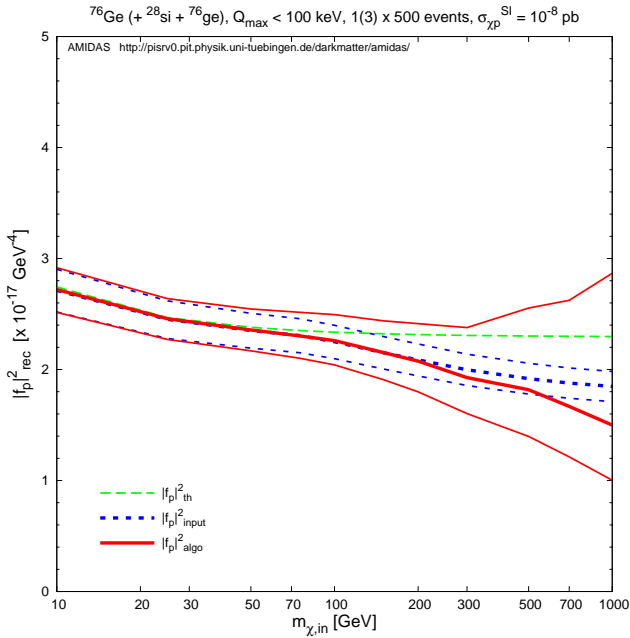}      \hspace{-1.1cm}
\includegraphics[width=9.8cm]{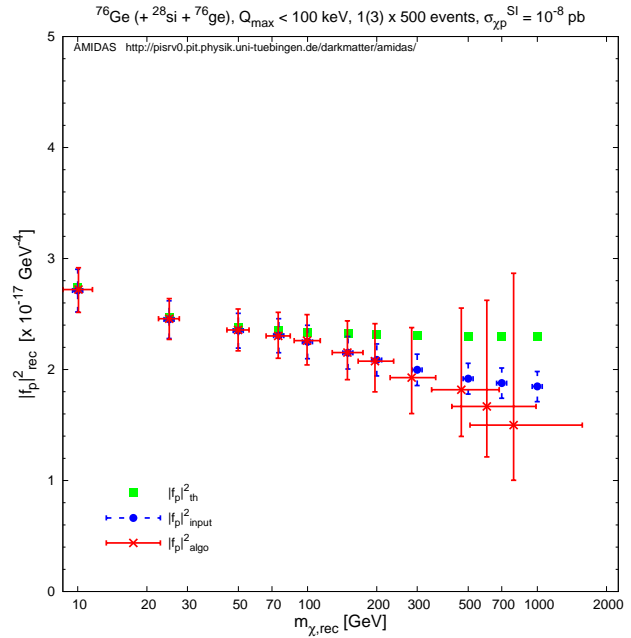} \hspace*{-1.6cm}
\vspace{-0.25cm}
\end{center}
\caption{
 As in Figs.~\ref{fig:fp2_rec} and \ref{fig:fp2_mchi_rec},
 except that
 the expected number of total events
 in {\em all three} experiments
 has been set as 500.
}
\label{fig:fp2_rec_500}
\end{figure}

 So far we have assumed that
 each experiment ``only'' has an exposure
 corresponding to 50 total events.
 In Figs.~\ref{fig:fp2_rec_500}
 we raise this number by a factor of 10.
 Not surprisingly,
 all uncertainties on both the reconstructed WIMP mass
 and the reconstructed SI couplings
 shrink by a factor $\gsim~3$
 compared to the results shown
 in Figs.~\ref{fig:fp2_mchi_rec}.
 Moreover,
 the small underestimate for lighter WIMP masses
 ($\mchi~\lsim$ 50 GeV)
 found in our simulations with only 50 events
 (see Figs.~\ref{fig:fp2_rec})
 disappears now.
 Note that,
 for heavier WIMP masses
 ($\mchi~\gsim$ 500 GeV),
 the upper bounds of the 1$\sigma$ statistical uncertainty
 on the reconstructed WIMP masses
 is now down to below our cut--off limit.
\section{Estimating the SD WIMP--nucleon couplings}
 For the sake of completeness,
 I consider briefly in this section the case
 that the spin--dependent WIMP--nucleus interaction
 dominates over the spin--independent one.
 Then the WIMP--nucleus cross section $\sigma_0$ in Eq.~(\ref{eqn:calA})
 can be expressed as \cite{SUSYDM96, Bertone05}:
\beq
   \sigmaSD
 = \afrac{32}{\pi} G_F^2 \~ \mrN^2
   \afrac{J + 1}{J} \bBig{\Srmp \armp + \Srmn \armn}^2
\~.
\label{eqn:sigma0SD}
\eeq
 Here $G_F$ is the Fermi constant,
 $J$ is the total spin of the target nucleus,
 $\expv{S_{\rm (p, n)}}$ are the expectation values of
 the proton and neutron group spins,
 and $a_{\rm (p, n)}$ are the effective SD WIMP couplings
 on protons and on neutrons.
 For the SD WIMP--nucleus cross section,
 it is usually assumed that
 only unpaired nucleons contribute significantly
 to the total cross section,
 as the spins of the nucleons in a nucleus
 are systematically anti--aligned%
\footnote{
 However,
 more detailed nuclear spin structure calculations show that
 the even group of nucleons has sometimes
 also non--negligible spin \cite{SUSYDM96}.
}.
 Under this assumption,
 the SD WIMP--nucleus cross section given above
 can be reduced to
\beqn
    \sigmaSD
 \= \afrac{32}{\pi} G_F^2 \~ \mrN^2
    \afrac{J + 1}{J} \expv{S_{\rm (p, n)}}^2 |a_{\rm (p, n)}|^2
    \non\\
 \= \frac{4}{3} \afrac{J + 1}{J} \expv{S_{\rm (p, n)}}^2
    \afrac{m_{\rm r, N}}{m_{\rm r, (p, n)}}^2
    \sigma_{\chi {\rm (p, n)}}^{\rm SD}
\~.
\label{eqn:sigma0SD_odd}
\eeqn
 Since
 for a proton or a neutron
 $J = \frac{1}{2}$ and $\Srmp$ or $\Srmn = \frac{1}{2}$,
 the SD WIMP cross section on protons or on neutrons
 can be given as
\beq
   \sigma_{\chi {\rm (p, n)}}^{\rm SD}
 = \afrac{24}{\pi} G_F^2 \~ m_{\rm r, (p, n)}^2 |a_{\rm (p, n)}|^2
\~.
\label{eqn:sigmap/nSD}
\eeq
 By comparing Eq.~(\ref{eqn:sigma0SD_odd})
 with the second expression in Eq.~(\ref{eqn:sigma0SI}),
 the {\em squared} SD WIMP couplings on protons and on neutrons
 can be obtained from Eq.~(\ref{eqn:fp2}) straightforwardly as
\beq
   |a_{\rm (p, n)}|^2
 = \frac{1}{\rho_0}
   \bbrac{\frac{\pi}{32 \sqrt{2}} \~ \Afrac{J}{J + 1} \!
          \afrac{1}{\calE G_F^2 \expv{S_{\rm (p, n)}}^2 \sqrt{\mN}}} \!\!
   \bbrac{\frac{2 \Qmin^{1/2} r(\Qmin)}{F_{\rm SD}^2(\Qmin)} + I_0}  \!
   \abrac{\mchi + \mN}
\~.
\label{eqn:ap/n2_odd}
\eeq
 Note that,
 for estimating $I_0$ here by using Eq.~(\ref{eqn:In_sum}),
 the elastic nuclear form factor $\FQ$
 must be chosen for the SD interaction.

 As the use of Eq.~(\ref{eqn:fp2}) for constraining
 the SI coupling $|f_{\rm p}|^2$ discussed in Sec.~2,
 by assuming that
 the SD WIMP--nucleus interaction contributes to
 the total cross section $\sigma_0$ dominantly,
 and combining with the conventional cross section--WIMP mass analysis,
 one could in principle use Eq.~(\ref{eqn:ap/n2_odd})
 to give the {\em lower} bounds of the WIMP mass
 and its SD couplings on protons and on neutrons
 from a {\em single} experiment with a target nucleus
 having spin sensitivity (almost) only on protons or on neutrons.
 By comparing these results with those obtained from the SI case,
 one could examine backwards
 the assumption for a dominant SD WIMP interaction.
 Meanwhile,
 as discussed in Sec.~3,
 once the WIMP mass $\mchi$ can be determined,
 one could then estimate the SD WIMP--nucleon cross sections
 by Eqs.~(\ref{eqn:ap/n2_odd}) and (\ref{eqn:sigmap/nSD})
 straightforwardly.

 Furthermore,
 by combining two target nuclei,
 one ($X$) of them has (almost) only spin sensitivity on protons
 and the other one ($Y$) on neutrons,
 one can easily find an expression
 for the {\em ratio} between two SD WIMP--nucleon couplings
 from Eq.~(\ref{eqn:ap/n2_odd}) as
\beqn
    \frac{\armn}{\armp}
 \= \pm
    \bbrac{\Afrac{J_Y}{J_Y + 1} \afrac{\calR_{\sigma, Y}}{\SnY^2}
           \afrac{\mchi + \mY}{\sqrt{\mY}}}^{1/2}
    \bbrac{\Afrac{J_X + 1}{J_X} \afrac{\SpX^2}{\calR_{\sigma, X}}
           \afrac{\sqrt{\mX}}{\mchi + \mX}}^{1/2}
    \non\\
 \= \pm \frac{\calR_{J, n, Y}}{\SnY} \cdot \frac{\SpX}{\calR_{J, n, X}}
\~.
\label{eqn:ranapSD_odd}
\eeqn
 Here I have used \cite{DMDDmchi-NJP}
\beq
        \calR_{\sigma, X}
 \equiv \frac{1}{\calE_X}
        \bbrac{\frac{2 \QminX^{1/2} r_X(\QminX)}{\FQminX} + \IzX}
\~,
\label{eqn:RsigmaX_min}
\eeq
 and defined
\beq
        \calR_{J, n, X}
 \equiv \bbrac{\Afrac{J_X}{J_X + 1}
               \frac{\calR_{\sigma, X}}{\calR_{n, X}}}^{1/2}
\label{eqn:RJnX}
\eeq
 for $n \neq 0$,
 with
\beqn
        \calR_{n, X}
 \equiv \bfrac{2 \QminX^{(n+1)/2} r_X(\QminX) / \FQminX + (n+1) \InX}
              {2 \QminX^{   1 /2} r_X(\QminX) / \FQminX +       \IzX}^{1/n}
\~;
\label{eqn:RnX_min}
\eeqn
 $\calR_{\sigma, Y}$, $\calR_{J, n, Y}$, and $\calR_{n, Y}$
 can be defined analogously.
 Here $m_{(X, Y)}$ and $F_{(X, Y)}(Q)$
 are the masses and the form factors of the nucleus $X$ and $Y$,
 respectively,
 $r_{(X, Y)}(Q_{{\rm min}, (X, Y)})$
 refer to the counting rates for the target $X$ and $Y$
 at the respective lowest recoil energies included in the analysis,
 and $\calE_{(X, Y)}$ are the experimental exposures
 with the target $X$ and $Y$.
 For the cancellation of the factors involving $\mchi$
 in the first line of Eq.~(\ref{eqn:ranapSD_odd}),
 I used the general estimator for the WIMP mass
 given in Refs.~\cite{DMDDmchi-SUSY07, DMDDmchi}:
\beq
   \left. \mchi \right|_{\Expv{v^n}}
 = \frac{\sqrt{\mX \mY} - \mX (\calR_{n, X} / \calR_{n, Y})}
        {\calR_{n, X} / \calR_{n, Y} - \sqrt{\mX / \mY}}
\~.
\label{eqn:mchi_Rn}
\eeq
 Note that
 Eq.~(\ref{eqn:ranapSD_odd}) derived here is in fact
 a special case of the general expression (16) given
 in Refs.~\cite{DMDDidentification-DARK2009,
                DMDDidentification-DMDE2009}.
 Detailed discussions
 about model--independent determinations of the ratios
 between different WIMP--nucleon couplings/cross sections
 can be found
 in Refs.~\cite{DMDDidentification-DARK2009,
                DMDDidentification-DMDE2009,
                DMDDranap}.
\section{Summary and conclusions}
 In this paper
 I presented the method for estimating
 the spin--independent WIMP--nucleon coupling
 from elastic WIMP--nucleus scattering experiments.
 This method is independent of the velocity distribution of halo WIMPs
 as well as (practically) of the as yet unknown WIMP mass.
 Assuming that
 an exponential--like shape of the recoil spectrum
 is confirmed from experimental data,
 the required information
 are only the measured recoil energies
 and the number of events in the first energy bin
 from at least two experiments with different target nuclei
 as well as the unique assumption for the local WIMP density.

 In Sec.~2
 I rederived the expression for estimating
 the (squared) SI WIMP--nucleon coupling $|f_{\rm p}|^2$
 as a function of the (unknown) WIMP mass
 \cite{DMDDidentification-DMDE2009}.
 Then I demonstrated that,
 by comparing the constrained area estimated by this method
 to that given by the conventional analysis
 with an assumed halo model,
 one could in principle
 -- for the first step with only one experiment
 observing positive signals --
 give the lower bounds of the WIMP mass
 and its SI cross section on nucleons
 from a single experiment.

 For the next step,
 I discussed in Sec.~3 that,
 by using measured recoil energies
 from two (or three) experiments with different target nuclei,
 we could not only determine the WIMP mass
 as discussed in Refs.~\cite{DMDDmchi-SUSY07, DMDDmchi},
 but also estimate the SI WIMP--nucleon coupling,
 with neither prior knowledge of each other
 nor an assumption for
 the velocity distribution of halo WIMPs.

 However,
 due to the degeneracy between
 the local WIMP density and the WIMP--nucleus cross section,
 it is impossible to determine both of them independently.
 As the simplest way
 one has thus to make an assumption for the local WIMP density.
 Nevertheless,
 since the SI WIMP--nucleon coupling
 is inversely proportional to the local WIMP density,
 whose common value
 would possibly be underestimated,
 one can then at least give an upper bound on this coupling.
 Moreover,
 our simulations show that,
 in spite of the very few ($\cal O$(50)) total events
 from one experiment,
 for a WIMP mass of 100 GeV,
 the SI WIMP--nucleon coupling can be estimated
 with a statistical uncertainty of only $\sim 15\%$;
 it leads to an uncertainty on the SI WIMP--nucleon cross section
 of only $\sim 30\%$,
 which is (much) smaller than the uncertainty
 on the estimate of the local Dark Matter density
 (of a factor of 2 or even larger).

 Our simulations show also that,
 due to (mainly) the experimental maximal cut--off energy,
 the SI WIMP coupling could be underestimated
 for heavier WIMP masses,
 especially with heavy target nuclei, e.g., Ge or Xe.
 However,
 since the kinematic maximum of recoil energies
 for heavier WIMP masses and/or with heavy target nuclei
 are (much) higher than for lighter WIMP masses
 with light nuclei,
 one could practically alleviate
 this systematic deviation
 by extending the detector sensitivity
 to higher energy ranges.
 Moreover,
 due to the fairly large statistical uncertainty,
 the true value of the SI WIMP--nucleon coupling lies always
 within the 1$\sigma$ statistical uncertainty interval.

 In Sec.~4
 I turned to consider the case that
 the spin--dependent WIMP--nucleus interaction
 dominates over the SI one.
 By assuming (naively) that
 only unpaired nucleons contribute significantly
 to the total WIMP--nucleus cross section,
 I gave also the expression for estimating
 the (squared) SD WIMP--nucleon couplings $|a_{\rm (p, n)}|^2$
 as functions of the (unknown) WIMP mass.
 As for the SI case,
 by comparing the constraints estimated by this method
 to those given by the conventional analysis,
 we could in principle also
 give the lower bounds of the WIMP mass
 and its SD cross sections on nucleons
 from a single experiment.

 Our simulations presented here are based on
 several simplified assumptions.
 Firstly,
 the sample to be analyzed contains only signal events,
 i.e., is free of background%
\footnote{
 For background discrimination techniques and status
 in currently running and projected direct detection experiments
 see e.g.,
 \cite{Aprile09a, CRESST-bg, EDELWEISS-bg, Ahmed09b}.
}$^{,}$
\footnote{
 For detailed simulations and discussions about
 effects of residue background events
 on the reconstructions of the WIMP mass
 and its SI coupling on nucleons see
 \cite{DMDDbg-mchi, DMDDbg-fp2}.
}.
 Secondly,
 all experimental systematic uncertainties as well as
 the uncertainty on the measurement of the recoil energy
 have been ignored.
 The energy resolution of most
 currently running and projected detectors
 is so good that its uncertainty can be neglected
 compared to the statistical uncertainty
 with (very) few events in the foreseeable future.

 A non--negligible threshold energy
 makes the conventional model--dependent analysis
 less sensitive on light WIMPs ($\mchi~\lsim$ 20 GeV),
 it could however give us more strict constraints
 on the lower bounds of the WIMP mass and its couplings on nucleons.
 In contrast,
 our simulation shows that,
 by using our model--independent method,
 the non--negligible threshold energy could cause
 not only a larger statistical uncertainty
 on the reconstructed couplings,
 but also a significant underestimate
 if WIMPs are (very) light.

 In summary,
 I demonstrated in this paper the use of our new method
 for estimating the spin--independent WIMP--nucleon coupling
 with neither a prior knowledge of the WIMP mass
 nor an assumption for the velocity distribution of halo WIMPs.
 By combining with information on
 the ratios between different WIMP--nucleon couplings/cross sections,
 which could also be determined model--independently
 \cite{DMDDidentification-DARK2009,
       DMDDidentification-DMDE2009,
       DMDDranap},
 one could in principle also estimate
 the absolute values of the spin--dependent cross sections.
 This information combined with the reconstructed WIMP mass
 will allow us not only to constrain the parameter space
 in different extensions of the Standard Model of particle physics
 \cite{Barger08, Belanger08, Cotta09},
 but also to identify WIMPs among new particles produced at colliders
 \cite{Baer}.
 Furthermore,
 knowledge of the WIMP mass and its couplings
 could not only offer a new approach for estimating
 the local WIMP density,
 but
 also
 permit the prediction of the WIMP annihilation cross section and
 the event rate in indirect Dark Matter detection experiments
 \cite{SUSYDM96, Bertone05}.
\subsubsection*{Acknowledgments}
 The author appreciates M.~Drees
 for useful discussions.
 The author would like to thank
 the Physikalisches Institut der Universit\"at T\"ubingen
 for the technical support of the computational work
 demonstrated in this article.
 This work
 was partially supported by
 the National Science Council of R.O.C.~%
 under contract no.~NSC-99-2811-M-006-031
 as well as by
 the LHC Physics Focus Group,
 National Center of Theoretical Sciences, R.O.C..
\appendix
\setcounter{equation}{0}
\setcounter{figure}{0}
\renewcommand{\theequation}{A\arabic{equation}}
\renewcommand{\thefigure}{A\arabic{figure}}
%
%
\section{Formulae needed in Secs.~2 and 3}
 Here I list all formulae needed
 for our model--independent data analyses
 described in this article.
 Detailed derivations and discussions
 can be found in Refs.~\cite{DMDDf1v, DMDDmchi}.
\subsection{Estimating \boldmath$r(\Qmin)$ and $I_n(\Qmin, \Qmax)$}
 First,
 consider experimental data described by
\beq
     {\T Q_n - \frac{b_n}{2}}
 \le \Qni
 \le {\T Q_n + \frac{b_n}{2}}
\~,
     ~~~~~~~~~~~~ 
     i
 =   1,~2,~\cdots,~N_n,~
     n
 =   1,~2,~\cdots,~B.
\label{eqn:Qni}
\eeq
 Here the total energy range between $\Qmin$ and $\Qmax$
 has been divided into $B$ bins
 with central points $Q_n$ and widths $b_n$.
 In each bin,
 $N_n$ events will be recorded.
 Since the recoil spectrum $dR / dQ$ is expected
 to be approximately exponential,
 the following ansatz for the {\em measured} recoil spectrum
 ({\em before} normalized by the experimental exposure $\calE$)
 in the $n$th bin has been introduced \cite{DMDDf1v}:
\beq
        \adRdQ_{{\rm expt}, \~ n}
 \equiv \adRdQ_{{\rm expt}, \~ Q \simeq Q_n}
 \equiv \rn  \~ e^{k_n (Q - Q_{s, n})}
\~.
\label{eqn:dRdQn}
\eeq
 Here $r_n$ is the standard estimator
 for $(dR / dQ)_{\rm expt}$ at $Q = Q_n$:
\beq
   r_n
 = \frac{N_n}{b_n}
\~,
\label{eqn:rn}
\eeq
 $k_n$ is the logarithmic slope of
 the recoil spectrum in the $n$th $Q-$bin,
 which can be computed numerically
 from the average value of the measured recoil energies
 in this bin:
\beq
   \bQn
 = \afrac{b_n}{2} \coth\afrac{k_n b_n}{2}-\frac{1}{k_n}
\~,
\label{eqn:bQn}
\eeq
 where
\beq
        \bQxn{\lambda}
 \equiv \frac{1}{N_n} \sumiNn \abrac{\Qni - Q_n}^{\lambda}
\~.
\label{eqn:bQn_lambda}
\eeq
 The error on the logarithmic slope $k_n$
 can be estimated from Eq.~(\ref{eqn:bQn}) directly as
\beq
   \sigma^2(k_n)
 = k_n^4
   \cbrac{  1
          - \bfrac{k_n b_n / 2}{\sinh (k_n b_n / 2)}^2}^{-2}
            \sigma^2\abrac{\bQn}
\~,
\label{eqn:sigma_kn}
\eeq
 with
\beq
   \sigma^2\abrac{\bQn}
 = \frac{1}{N_n - 1} \bbigg{\bQQn - \bQn^2}
\~.
\label{eqn:sigma_bQn}
\eeq
 $Q_{s, n}$ in the ansatz (\ref{eqn:dRdQn})
 is the shifted point at which
 the leading systematic error due to the ansatz
 is minimal \cite{DMDDf1v},
\beq
   Q_{s, n}
 = Q_n + \frac{1}{k_n} \ln\bfrac{\sinh(k_n b_n/2)}{k_n b_n/2}
\~.
\label{eqn:Qsn}
\eeq
 Note that $Q_{s, n}$ differs from
 the central point of the $n$th bin, $Q_n$.
 From the ansatz (\ref{eqn:dRdQn}),
 the counting rate at $Q = \Qmin$ can be calculated by
\beq
   r(\Qmin)
 = r_1 e^{k_1 (\Qmin - Q_{s, 1})}
\~,
\label{eqn:rmin_eq}
\eeq
 and its statistical error can be expressed as
\beq
   \sigma^2(r(\Qmin))
 = r^2(\Qmin) 
   \cbrac{  \frac{1}{N_1}
          + \bbrac{  \frac{1}{k_1}
                   - \afrac{b_1}{2} 
                     \abrac{  1
                            + \coth\afrac{b_1 k_1}{2}}}^2
            \sigma^2(k_1)}
\~,
\label{eqn:sigma_rmin}
\eeq
 since
\beq
   \sigma^2(r_n)
 = \frac{N_n}{b_n^2}
\~.
\label{eqn:sigma_rn}
\eeq
 Finally,
 since all $I_n$ are determined from the same data,
 they are correlated with
\beq
   {\rm cov}(I_n, I_m)
 = \sum_{a = 1}^{N_{\rm tot}} \frac{Q_a^{(n+m-2)/2}}{F^4(Q_a)}
\~,
\label{eqn:cov_In}
\eeq
 where the sum runs over all events
 with recoil energy between $\Qmin$ and $\Qmax$.
 And the correlation between the errors on $r(\Qmin)$,
 which is calculated entirely
 from the events in the first bin,
 and on $I_n$ is given by
\beqn
 \conti {\rm cov}(r(\Qmin), I_n)
        \non\\
 \=     r(\Qmin) \~ I_n(\Qmin, \Qmin + b_1)
        \non\\
 \conti ~~~~ \times 
        \cleft{  \frac{1}{N_1} 
               + \bbrac{  \frac{1}{k_1}
                        - \afrac{b_1}{2} \abrac{1 + \coth\afrac{b_1 k_1}{2}}}}
        \non\\
 \conti ~~~~~~~~~~~~~~ \times 
        \cright{ \bbrac{  \frac{I_{n+2}(\Qmin, \Qmin + b_1)}
                               {I_{n  }(\Qmin, \Qmin + b_1)}
                        - Q_1
                        + \frac{1}{k_1}
                        - \afrac{b_1}{2} \coth\afrac{b_1 k_1} {2}}
        \sigma^2(k_1)}
\~;
\label{eqn:cov_rmin_In}
\eeqn
 note that
 the sums $I_i$ here only count in the first bin,
 which ends at $Q = \Qmin + b_1$.

 On the other hand,
 with a functional form of the recoil spectrum
 (e.g., fitted to experimental data),
 $(dR / dQ)_{\rm expt}$,
 one can use the following integral forms
 to replace the summations given above.
 Firstly,
 the average $Q-$value in the $n$th bin
 defined in Eq.~(\ref{eqn:bQn_lambda})
 can be calculated by
\beq
   \bQxn{\lambda}
 = \frac{1}{N_n} \intQnbn \abrac{Q - Q_n}^{\lambda} \adRdQ_{\rm expt} dQ
\~.
\label{eqn:bQn_lambda_int}
\eeq
 For $I_n(\Qmin, \Qmax)$ given in Eq.~(\ref{eqn:In_sum}),
 we have
\beq
   I_n(\Qmin, \Qmax)
 = \int_{\Qmin}^{\Qmax} \frac{Q^{(n-1)/2}}{F^2(Q)} \adRdQ_{\rm expt} dQ
\~,
\label{eqn:In_int}
\eeq 
 and similarly for the covariance matrix for $I_n$
 in Eq.~(\ref{eqn:cov_In}),
\beq
   {\rm cov}(I_n, I_m)
 = \int_{\Qmin}^{\Qmax} \frac{Q^{(n+m-2)/2}}{F^4(Q)} \adRdQ_{\rm expt} dQ
\~.
\label{eqn:cov_In_int}
\eeq 
 Remind that
 $(dR / dQ)_{\rm expt}$ is the {\em measured} recoil spectrum
 {\em before} normalized by the exposure.
 Finally,
 $I_i(\Qmin, \Qmin + b_1)$ needed in Eq.~(\ref{eqn:cov_rmin_In})
 can be calculated by
\beq
   I_n(\Qmin, \Qmin + b_1)
 = \int_{\Qmin}^{\Qmin + b_1}
   \frac{Q^{(n-1)/2}}{F^2(Q)} \bbigg{r_1 \~ e^{k_1 (Q - Q_{s, 1})}} dQ
\~.
\label{eqn:In_1_int}
\eeq 
 Note that,
 firstly,
 $r(\Qmin)$ and $I_n(\Qmin, \Qmin + b_1)$ should be
 estimated by Eqs.~(\ref{eqn:rmin_eq}) and (\ref{eqn:In_1_int})
 with $r_1$, $k_1$ and $Q_{s, 1}$
 estimated by Eqs.~(\ref{eqn:rn}), (\ref{eqn:bQn}), and (\ref{eqn:Qsn})
 in order to use the other formulae for estimating
 the (correlations between the) statistical errors
 without any modification.
 Secondly,
 $r(\Qmin)$ and $I_n(\Qmin, \Qmax)$ estimated
 from a scattering spectrum fitted to experimental data
 are usually not model--independent any more.
 Moreover,
 for estimating the SD WIMP--nucleon couplings
 by Eq.~(\ref{eqn:ap/n2_odd}),
 the elastic nuclear form factor $\FQ$
 in Eqs.~(\ref{eqn:In_sum}), (\ref{eqn:cov_In}), (\ref{eqn:In_int}),
 (\ref{eqn:cov_In_int}), and (\ref{eqn:In_1_int})
 should be understood to be chosen
 for the SD interaction.
\subsection{Determining the WIMP mass \boldmath$\mchi$}
 By requiring that
 the values of a given moment of $f_1(v)$
 estimated by Eq.~(\ref{eqn:moments})
 from two experiments
 with different target nuclei, $X$ and $Y$, agree,
 $\mchi$ appearing in the prefactor $\alpha^n$
 on the right--hand side of Eq.~(\ref{eqn:moments})
 can be solved analytically as
 \cite{DMDDmchi-SUSY07, DMDDmchi}:
\cheqnref{eqn:mchi_Rn}
\beq
   \left. \mchi \right|_{\Expv{v^n}}
 = \frac{\sqrt{\mX \mY} - \mX (\calR_{n, X} / \calR_{n, Y})}
        {\calR_{n, X} / \calR_{n, Y} - \sqrt{\mX / \mY}}
\~,
\eeq
\cheqnXN{A}{-1}%
 with $\calR_{n, (X, Y)}$ given by Eq.~(\ref{eqn:RnX_min}).
 Note that
 the general expression (\ref{eqn:mchi_Rn}) can be used
 either for spin--independent or
 for spin--dependent scattering,
 one only needs to choose different form factors
 under different assumptions;
 the form factors needed for estimating $I_{n, (X, Y)}$
 by Eq.~(\ref{eqn:In_sum}) or (\ref{eqn:In_int})
 are thus also different.

 By using the standard Gaussian error propagation,
 a lengthy expression for the statistical uncertainty on
 $\left. \mchi \right|_{\Expv{v^n}}$
 can be obtained as
\beqn
        \left. \sigma(\mchi) \right|_{\Expv{v^n}}
 \=     \frac{\sqrt{\mX / \mY} \vbrac{\mX - \mY} \abrac{\calR_{n, X} / \calR_{n, Y}} }
             {\abrac{\calR_{n, X} / \calR_{n, Y} - \sqrt{\mX / \mY}}^2}
        \non\\
 \conti ~~~~ \times
        \bbrac{  \frac{1}{\calR_{n, X}^2}
                 \sum_{i, j = 1}^3
                 \aPp{\calR_{n, X}}{c_{i, X}} \aPp{\calR_{n, X}}{c_{j, X}}
                 {\rm cov}(c_{i, X}, c_{j, X})
               + (X \lto Y)}^{1 / 2}
\!.
\label{eqn:sigma_mchi_Rn}
\eeqn
 Here a short--hand notation for the six quantities
 on which the estimate of $\mchi$ depends
 has been introduced:
\beq
   c_{1, X}
 = I_{n, X}
\~,
   ~~~~~~~~~~~~ 
   c_{2, X}
 = I_{0, X}
\~,
   ~~~~~~~~~~~~ 
   c_{3, X}
 = r_X(\QminX)
\~;
\label{eqn:ciX}
\eeq
 and similarly for the $c_{i, Y}$.
 Estimators for ${\rm cov}(c_i, c_j)$ have been given
 in Eqs.~(\ref{eqn:cov_In}) and (\ref{eqn:cov_rmin_In}).
 Explicit expressions for the derivatives of $\calR_{n, X}$
 with respect to $c_{i, X}$ are:
\cheqnXa{A}
\beq
   \Pp{\calR_{n, X}}{\InX}
 = \frac{n + 1}{n}
   \bfrac{\FQminX}{2 \QminX^{(n + 1) / 2} r_X(\QminX) + (n + 1) \InX \FQminX}
   \calR_{n, X}
\~,
\label{eqn:dRnX_dInX}
\eeq
\cheqnXb{A}
\beq
   \Pp{\calR_{n, X}}{\IzX}
 =-\frac{1}{n}
   \bfrac{\FQminX}{2 \QminX^{1 / 2} r_X(\QminX) + \IzX \FQminX}
   \calR_{n, X}
\~,
\label{eqn:dRnX_dIzX}
\eeq
 and
\cheqnXc{A}
\beqn
        \Pp{\calR_{n, X}}{r_X(\QminX)}
 \=     \frac{2}{n}
        \bfrac{  \QminX^{(n + 1) / 2} \IzX        - (n + 1) \QminX^{1 / 2} \InX}
              {2 \QminX^{(n + 1) / 2} r_X(\QminX) + (n + 1) \InX \FQminX}
        \non\\
 \conti ~~~~~~~~~~~~~~~~ \times 
        \bfrac{\FQminX}{2 \QminX^{1 / 2} r_X(\QminX) + \IzX \FQminX}
        \calR_{n, X}
\~;
\label{eqn:dRnX_drminX}
\eeqn
\cheqnX{A}%
 explicit expressions for the derivatives of $\calR_{n, Y}$
 with respect to $c_{i, Y}$ can be given analogously.
 Note that,
 firstly,
 factors $\calR_{n, (X, Y)}$ appear in all these expressions,
 which can practically be cancelled by the prefactors
 in the bracket in Eq.~(\ref{eqn:sigma_mchi_Rn}).
 Secondly,
 all the $I_{0, (X, Y)}$ and $I_{n, (X, Y)}$ should be understood
 to be computed according to
 Eq.~(\ref{eqn:In_sum}) or (\ref{eqn:In_int})
 with integration limits $\Qmin$ and $\Qmax$
 specific for that target.

 On the other hand,
 since $|f_{\rm p}|^2$ in Eq.~(\ref{eqn:fp2})
 is identical for different targets,
 it leads to a second expression for determining $\mchi$
 \cite{DMDDmchi}:
\beq
   \left. \mchi \right|_\sigma
 = \frac{\abrac{\mX / \mY}^{5 / 2} \mY - \mX (\calR_{\sigma, X} / \calR_{\sigma, Y})}
        {\calR_{\sigma, X} / \calR_{\sigma, Y} - \abrac{\mX / \mY}^{5 / 2}}
\~.
\label{eqn:mchi_Rsigma}
\eeq
 Here $m_{(X, Y)} \propto A_{(X, Y)}$ has been assumed,
 and $\calR_{\sigma, (X, Y)}$ have been given
 in Eq.~(\ref{eqn:RsigmaX_min}).
 Similar to the analogy between
 Eqs.~(\ref{eqn:mchi_Rn}) and (\ref{eqn:mchi_Rsigma}),
 the statistical uncertainty on $\left. \mchi \right|_\sigma$
 can be expressed as
\beqn
        \left. \sigma(\mchi) \right|_\sigma
 \=     \frac{\abrac{\mX / \mY}^{5 / 2} \vbrac{\mX - \mY}
              \abrac{\calR_{\sigma, X} / \calR_{\sigma, Y}} }
             {\bbrac{\calR_{\sigma, X} / \calR_{\sigma, Y} - \abrac{\mX / \mY}^{5 / 2}}^2}
        \non\\
 \conti ~~~~~~ \times 
        \bbrac{  \frac{1}{\calR_{\sigma, X}^2}
                 \sum_{i, j = 2}^3
                 \aPp{\calR_{\sigma, X}}{c_{i, X}} \aPp{\calR_{\sigma, X}}{c_{j, X}}
                 {\rm cov}(c_{i, X}, c_{j, X})
               + (X \lto Y)}^{1 / 2}
\~,
\label{eqn:sigma_mchi_Rsigma}
\eeqn
 where I have used again
 the short--hand notation in Eq.~(\ref{eqn:ciX});
 note that $c_{1, (X, Y)} = I_{n, (X, Y)}$ do not appear here.
 Expressions for the derivatives of $\calR_{\sigma, X}$
 can be computed from Eq.~(\ref{eqn:RsigmaX_min}) as
\cheqnXa{A}
\beq
   \Pp{\calR_{\sigma, X}}{\IzX}
 = \bfrac{\FQminX}{2 \QminX^{1 / 2} r_X(\QminX) + \IzX \FQminX}
   \calR_{\sigma, X}
\~,
\label{eqn:dRsigmaX_dIzX}
\eeq
\cheqnXb{A}
\beq
   \Pp{\calR_{\sigma, X}}{r_X(\QminX)}
 = \bfrac{2 \QminX^{1 / 2}}{2 \QminX^{1 / 2} r_X(\QminX) + \IzX \FQminX}
   \calR_{\sigma, X}
\~;
\label{eqn:dRsigmaX_drminX}
\eeq
\cheqnX{A}%
 and similarly for the derivatives of $\calR_{\sigma, Y}$.
 Remind that
 factors $\calR_{\sigma, (X, Y)}$ appearing here
 can also be cancelled by the prefactors
 in the bracket in Eq.~(\ref{eqn:sigma_mchi_Rsigma}).

 In order to yield the best--fit WIMP mass
 as well as to minimize its statistical uncertainty
 by combining the estimators for different $n$
 in Eq.~(\ref{eqn:mchi_Rn}) with each other
 and with the estimator in Eq.~(\ref{eqn:mchi_Rsigma}),
 a $\chi^2$ function has been introduced as
 \cite{DMDDmchi}
\beq
   \chi^2(\mchi)
 = \sum_{i, j}
   \abrac{f_{i, X} - f_{i, Y}} {\cal C}^{-1}_{ij} \abrac{f_{j, X} - f_{j, Y}}
\~,
\label{eqn:chi2}
\eeq
 where
\cheqnXa{A}
\beqn
           f_{i, X}
 \eqnequiv \alpha_X^i
           \bfrac{  2 Q_{{\rm min}, X}^{(i+1)/2} r_X(\Qmin) / F^2_X(Q_{{\rm min}, X})
                  + (i+1) I_{i, X}}
                 {  2 Q_{{\rm min}, X}^{   1 /2} r_X(\Qmin) / F^2_X(Q_{{\rm min}, X})
                  +       \IzX}
           \afrac{1}{300~{\rm km/s}}^i
           \non\\
           \non\\
 \=        \afrac{\alpha_X {\cal R}_{i, X}}{300~{\rm km/s}}^{i}
\~,
\label{eqn:fiXa}
\eeqn
 for $i = -1,~1,~2,~\dots,~n_{\rm max}$, and
\cheqnXb{A}
\beqn
           f_{n_{\rm max}+1, X}
 \eqnequiv \calE_X
           \bfrac{A_X^2}
                 {  2 Q_{{\rm min}, X}^{1/2} r_X(\Qmin) / F^2_X(Q_{{\rm min}, X})
                  + \IzX}
           \afrac{\sqrt{\mX}}{\mchi + \mX}
           \non\\
           \non\\
 \=        \frac{A_X^2}{\calR_{\sigma, X}} \afrac{\sqrt{\mX}}{\mchi + \mX}
\~;
\label{eqn:fiXb}
\eeqn
\cheqnX{A}%
 the other $n_{\rm max} + 2$ functions $f_{i, Y}$
 can be defined analogously.
 Here $n_{\rm max}$ determines the highest moment of $f_1(v)$
 that is included in the fit.
 The $f_i$ are normalized such that
 they are dimensionless and very roughly of order unity
 in order to alleviate numerical problems
 associated with the inversion of their covariance matrix.
 Note that
 the first $n_{\rm max} + 1$ fit functions
 depend on $\mchi$ only through the overall factor $\alpha$
 and $\mchi$ in Eqs.~(\ref{eqn:fiXa}) and (\ref{eqn:fiXb})
 is now a fit parameter,
 which may differ from the true value of the WIMP mass.
 Finally,
 $\cal C$ in Eq.~(\ref{eqn:chi2}) is the total covariance matrix.
 Since the $X$ and $Y$ quantities
 are statistically completely independent,
 $\cal C$ can be written as a sum of two terms:
\beq
   {\cal C}_{ij}
 = {\rm cov}\abrac{f_{i, X}, f_{j, X}} + {\rm cov}\abrac{f_{i, Y}, f_{j, Y}}
\~.
\label{eqn:Cij}
\eeq
 The entries of the $\cal C$ matrix given here
 involving basically only
 the moments of the WIMP velocity distribution
 can be read off Eq.~(82) of Ref.~\cite{DMDDf1v},
 with an slight modification
 due to the normalization factor in Eq.~(\ref{eqn:fiXa})%
\footnote{
 Since the last $f_i$ defined in Eq.~(\ref{eqn:fiXb})
 can be computed from the same basic quantities,
 i.e., the counting rates at $\Qmin$ and the integrals $I_0$,
 it can directly be included in the covariance matrix.
}:
\beqn
        {\rm cov}\abrac{f_i, f_j}
 \=     \calN_{\rm m}^2
        \bbiggl{  f_i \~ f_j \~ {\rm cov}(I_0, I_0)
                + \Td{\alpha}^{i+j} (i+1) (j+1) {\rm cov}(I_i, I_j)}
        \non\\
 \conti ~~~~~~~~ 
                - \Td{\alpha}^j (j+1) f_i \~ {\rm cov}(I_0, I_j)
                - \Td{\alpha}^i (i+1) f_j \~ {\rm cov}(I_0, I_i)\bigg.
        \non\\
 \conti ~~~~~~~~~~~~ 
                + D_i D_j \sigma^2(r(\Qmin))
                - \abrac{D_i f_j + D_j f_i} {\rm cov}(r(\Qmin), I_0)\Bigg.
        \non\\
 \conti ~~~~~~~~~~~~~~~~ 
        \bbiggr{+ \Td{\alpha}^j (j+1) D_i \~ {\rm cov}(r(\Qmin), I_j)
                + \Td{\alpha}^i (i+1) D_j \~ {\rm cov}(r(\Qmin), I_i)}
\~.
        \non\\
\label{eqn:cov_fi}
\eeqn
 Here I used
\cheqnref{eqn:calNm}
\beq
        \calN_{\rm m}
 \equiv \frac{1}{2 \Qmin^{1 /2} r(\Qmin)/\FQmin + I_0}
\~,
\eeq
\cheqnXN{A}{-1}%
\beq
        \Td{\alpha}
 \equiv \frac{\alpha}{300~{\rm km/s}}
\~,
\label{eqn:td_alpha}
\eeq
 and
\cheqnXa{A}
\beq
        D_i
 \equiv \frac{1}{\cal N_{\rm m}} \bPp{f_i}{r(\Qmin)}
 =      \frac{2}{\FQmin}
        \abigg{\Td{\alpha}^i \Qmin^{(i+1)/2} - \Qmin^{1/2} \~ f_i}
\~,
\label{eqn:Dia}
\eeq
 for $i = -1,~1,~2,~\dots,~n_{\rm max}$; and
\cheqnXb{A}
\beq
   D_{n_{\rm max}+1}
 = \frac{2}{\FQmin} \abrac{-\Qmin^{1/2} f_{n_{\rm max}+1}}
\~.
\label{eqn:Dib}
\eeq
\cheqnX{A}%

 Finally,
 since the basic requirement of the expressions for determining $\mchi$
 given in Eqs.~(\ref{eqn:mchi_Rn}) and (\ref{eqn:mchi_Rsigma}) is that,
 from two experiments with different target nuclei,
 the values of a given moment of the WIMP velocity distribution
 estimated by Eq.~(\ref{eqn:moments}) should agree,
 the upper cuts on $f_1(v)$ in two data sets
 should be (approximately) equal%
\footnote{
 Here the threshold energies
 have been assumed to be negligible.
}.
 Since $v_{\rm cut} = \alpha \sqrt{Q_{\rm max}}$,
 it requires that \cite{DMDDmchi}
\beq
   Q_{{\rm max}, Y}
 = \afrac{\alpha_X}{\alpha_Y}^2 Q_{{\rm max}, X}
\~.
\label{eqn:match}  
\eeq
 Note that
 $\alpha$ defined in Eq.~(\ref{eqn:alpha})
 is a function of the true WIMP mass.
 Thus this relation for matching optimal cut--off energies
 can be used only if $\mchi$ is already known.
 One possibility to overcome this problem is
 to fix the cut--off energy of the experiment with the heavier target,
 minimize the $\chi^2(\mchi)$ function
 defined in Eq.~(\ref{eqn:chi2}),
 and then estimate the cut--off energy for the lighter nucleus
 by Eq.~(\ref{eqn:match}) algorithmically \cite{DMDDmchi}.
\subsection{Covariance of \boldmath$\mchi$ and $1 / \calN_{\rm m}$}
 First,
 the statistical error on $1 / \calN_{\rm m}$
 can be given from Eq.~(\ref{eqn:calNm}) directly as
\beq
      \sigma^2(1 / \calN_{\rm m})
 =    \bfrac{2 \Qmin^{1 / 2}}{\FQmin}^2 \sigma^2(r(\Qmin))
  +   \sigma^2(I_0)
  + 2 \bfrac{2 \Qmin^{1 / 2}}{\FQmin} {\rm cov}(r(\Qmin), I_0)
\~.
\label{eqn:sigma2_calNm}
\eeq
 For the case that
 one has only two data sets with different target nuclei, $X$ and $Y$,
 one of these two data sets will then be needed
 for reconstructing the WIMP mass $\mchi$ and
 also for estimating $1 / \calN_{\rm m}$ in Eq.~(\ref{eqn:fp2}).
 The uncertainties on $\mchi$ and $1 / \calN_{\rm m}$
 are thus correlated.
 Assuming that
 the WIMP mass is reconstructed by Eq.~(\ref{eqn:mchi_Rn}),
 and target $X (Y)$ is used for estimating $1 / \calN_{\rm m}$,
 the covariance of $\left. \mchi \right|_{\Expv{v^n}}$
 and $1 / \calN_{{\rm m}, (X, Y)}$ can be obtained
 by modifying Eq.~(\ref{eqn:sigma_mchi_Rn}) slightly as
\cheqnXa{A}
\beqn
 \conti {\rm cov}(\left. \mchi \right|_{\Expv{v^n}}, 1 / \calN_{{\rm m}, X})
        \non\\
 \=     \frac{\sqrt{\mX / \mY} \abrac{\mX - \mY} \abrac{\calR_{n, X} / \calR_{n, Y}} }
             {\abrac{\calR_{n, X} / \calR_{n, Y} - \sqrt{\mX / \mY}}^2}
        \afrac{1}{\calR_{n, X}}
        \non\\
 \conti ~~~~~~ \times 
               \sum_{i = 1}^3
               \aPp{\calR_{n, X}}{c_{i, X}}
               \bbrac{  {\rm cov}(c_{i, X}, \IzX)
                      + {\rm cov}(c_{i, X}, r_X(\QminX))
                        \afrac{2 \QminX^{1 / 2}}{\FQminX} }
\~,
\eeqn
 and
\cheqnXb{A}
\beqn
 \conti {\rm cov}(\left. \mchi \right|_{\Expv{v^n}}, 1 / \calN_{{\rm m}, Y})
        \non\\
 \=     \frac{\sqrt{\mX / \mY} \abrac{\mX - \mY} \abrac{\calR_{n, X} / \calR_{n, Y}} }
             {\abrac{\calR_{n, X} / \calR_{n, Y} - \sqrt{\mX / \mY}}^2}
        \afrac{-1}{\calR_{n, Y}}
        \non\\
 \conti ~~~~~~ \times 
               \sum_{i = 1}^3
               \aPp{\calR_{n, Y}}{c_{i, Y}}
               \bbrac{  {\rm cov}(c_{i, Y}, \IzY)
                      + {\rm cov}(c_{i, Y}, r_Y(\QminY))
                        \afrac{2 \QminY^{1 / 2}}{\FQminY} }
\~.
\eeqn
\cheqnX{A}%
 For the case that
 the WIMP mass is reconstructed by Eq.~(\ref{eqn:mchi_Rsigma}),
 one can also modify Eq.~(\ref{eqn:sigma_mchi_Rsigma})
 to obtain that
\cheqnXa{A}
\beqn
 \conti {\rm cov}(\left. \mchi \right|_\sigma, 1 / \calN_{{\rm m}, X})
        \non\\
 \=     \frac{\abrac{\mX / \mY}^{5 / 2} \abrac{\mX - \mY}
              \abrac{\calR_{\sigma, X} / \calR_{\sigma, Y}} }
             {\bbrac{\calR_{\sigma, X} / \calR_{\sigma, Y} - \abrac{\mX / \mY}^{5 / 2}}^2}
        \afrac{1}{\calR_{\sigma, X}}
        \non\\
 \conti ~~~~~~ \times 
               \sum_{i = 2}^3
               \aPp{\calR_{\sigma, X}}{c_{i, X}}
               \bbrac{  {\rm cov}(c_{i, X}, \IzX)
                      + {\rm cov}(c_{i, X}, r_X(\QminX))
                        \afrac{2 \QminX^{1 / 2}}{\FQminX} }
\~,
\eeqn
 and
\cheqnXb{A}
\beqn
 \conti {\rm cov}(\left. \mchi \right|_\sigma, 1 / \calN_{{\rm m}, Y})
        \non\\
 \=     \frac{\abrac{\mX / \mY}^{5 / 2} \abrac{\mX - \mY}
              \abrac{\calR_{\sigma, X} / \calR_{\sigma, Y}} }
             {\bbrac{\calR_{\sigma, X} / \calR_{\sigma, Y} - \abrac{\mX / \mY}^{5 / 2}}^2}
        \afrac{-1}{\calR_{\sigma, Y}}
        \non\\
 \conti ~~~~~~ \times 
               \sum_{i = 2}^3
               \aPp{\calR_{\sigma, Y}}{c_{i, Y}}
               \bbrac{  {\rm cov}(c_{i, Y}, \IzY)
                      + {\rm cov}(c_{i, Y}, r_Y(\QminY))
                        \afrac{2 \QminY^{1 / 2}}{\FQminY} }
\~.
\eeqn
\cheqnX{A}%
 Note that,
 firstly,
 in the above expressions
 we have to use $\abrac{\mX - \mY}$
 instead of $\vbrac{\mX - \mY}$
 in Eqs.~(\ref{eqn:sigma_mchi_Rn}) and (\ref{eqn:sigma_mchi_Rsigma});
 for expressions with the $Y$ target,
 there is an additional ``$-$ (minus)'' sign.
 Secondly,
 the algorithmic process for matching
 the experimental maximal cut--off energies of two experiments
 used for the reconstruction of the WIMP mass
 can also be used with the basic expressions
 (\ref{eqn:mchi_Rn}) and (\ref{eqn:mchi_Rsigma}).
 For this case and the lighter nucleus
 is used for estimating $1 / \calN_{\rm m}$,
 the energy range
 of the sum in Eq.~(\ref{eqn:cov_In}) or
 of the integral in Eq.~(\ref{eqn:cov_In_int})
 as the estimator for the covariance of $I_n$
 should be modified to be between $\Qmin$ and
 the {\em reduced} maximal cut--off energy
 of the lighter nucleus.
\end{document}